\documentclass[iop, twocolappendix]{emulateapj}

\usepackage{epstopdf}
\usepackage{graphicx}
\usepackage{natbib}
\usepackage{amsmath}

\shorttitle{The GENGA Code}
\shortauthors{Grimm and Stadel}

\begin{document}

\title{The GENGA Code: Gravitational Encounters in $N$-body simulations with GPU Acceleration}

\author{Simon L. Grimm \& Joachim G. Stadel }
\affil{Institute for Computational Science, University of Z\"{u}rich}
\affil{Winterthurerstrasse 190, CH-8057, Z\"{u}rich, Switzerland}
\email{sigrimm@physik.uzh.ch}

\begin{abstract}
We describe an open source GPU implementation of a hybrid symplectic $N$-body integrator, GENGA ({\bf G}ravitational {\bf EN}counters with {\bf G}pu {\bf A}cceleration), designed to integrate planet and planetesimal dynamics in the late stage of planet formation and stability analyses of planetary systems. GENGA uses a hybrid symplectic integrator to handle close encounters with very good energy conservation, which is essential in long-term planetary system integration. We extended the second order hybrid integration scheme to higher orders. The GENGA code supports three simulation modes: Integration of up to 2048 massive bodies, integration with up to a million test particles, or parallel integration of a large number of individual planetary systems. We compare the results of GENGA to Mercury and pkdgrav2 in respect of energy conservation and performance, and find that the energy conservation of GENGA is comparable to Mercury and around two orders of magnitude better than pkdgrav2. GENGA runs up to 30 times faster than Mercury and up to eight times faster than pkdgrav2.  
GENGA is written in CUDA C and runs on all NVIDIA GPUs with compute capability of at least 2.0.

\end{abstract}


\section{Introduction}

The use of numerical $N$-body simulations to study the evolution of gravitational many-body systems, in particular that of our solar system, has a long tradition in astronomy. Prediction of planetary positions have occupied the field since the time of Newton, while at present the study of the full physical process of the formation and evolution of planetary systems requires a very significant amount of computing resources. We present here a new $N$-body integrator, GENGA\footnote{GENGA is available as open source code from https://bitbucket.org/sigrimm/genga} ({\bf G}ravitational {\bf EN}counters with {\bf G}pu {\bf A}cceleration), which uses today's most efficient computing hardware: the graphical processing units (GPUs). GENGA supports three computing modes: simulations of up to 2048 planetesimals, simulations with up to a million massless test particles, and parallel simulations of a large number of small planetary systems.

\subsection{Physical Motivation}
In the following we give a short overview of past work done in planetary $N$-body simulations which has motivated and guided the development of GENGA. This has mainly focused on four application areas, namely long-term evolution and stability analysis of the solar system, the dynamics of small asteroids under the gravitational influence of the planets, the planet formation process with the dynamics of planetesimals, and finally the evolution and characterization of exoplanetary systems.

The long-term evolution and stability analysis of the solar system using $N$-body integrations has been studied by several people. A first result on an instability of the solar system was found by \cite{SussmannWisdom1988}, which integrated the five outer planets (including Pluto) over 845 Myr and found a Lyapunov time of Pluto's motion of 20 Myr. Including also the inner planets into the integration is more challenging because a much smaller time step is needed for a comparable accuracy.
The entire solar system with all nine planets (including Pluto) and the Earth moon was performed by \cite{Quinn91} over 3 Myr backward in time. A longer simulation over 98.6 Myr was performed by \cite{SussmaWisdom92}, which used a symplectic $N$-body mapping \citep{WisdomHolman91, Gladman+1990, SahaTremaine1992} and confirmed the previous result on the chaotic motion of Pluto. They also confirmed the Lyapunov time of 5 Myr of the solar system, predicted by \cite{Laskar1989}. A small perturbation in the initial conditions of one of the inner planets can have dramatic effects on the evolution of the solar system. Even collisions between planets are possible in less than 3.5 Gyr \citep{Laskar96}. An overview of the question of stability of the solar system can be found in \cite{Laskar2012}.

In addition to the massive planets in the solar system, massless test particles can be used to study the dynamics of meteorites, comets or impact ejecta. Since test particles do not interact with other test particles, the number of interactions that need to be calculated is greatly reduced.
\cite{Gladman+1996} simulated 2100 particles escaping from Mars due to an impact and found a delivery efficiency to Earth of $7.5\%$ for $v_\infty = 1 \mathtt{kms^{-1}}$. They simulated also 200 particles escaping from Mercury and found one particle hitting the Earth after 23 Myr. The trajectories from Earth impact ejecta are studied by \cite{Wells+2003} with the PKDGRAV code \citep{Stadel2011} and they found 9 out of 675 particles returning to the Earth after 3000-5000 yr. They concluded that micro-biological ejecta could survive a sterilizing impact and reseed the Earth again with life.
The delivery rates of terrestrial material to Mars and Venus and also back to Earth was studied by \cite{Gladman+2005}, and the trajectories of Mercurial material in more detail, by \cite{GladmanCoffey2009}.
\cite{ReyesRuiz+2012} simulated $10^5$ test particles for 30,000 yr and found Earth ejecta reaching the Moon, Venus, Mars, Jupiter and Saturn, which means that biological material could in principle be transferred to other planets or their satellites by an impact to the Earth.

To simulate the late stage of planet formation, starting from the runaway growth phase when planetesimals collide and form bigger objects like planetary embryos, test particles cannot be used anymore, and all $N^2$ gravitational interactions between the planetesimals have to be taken into account. Also close encounters between planetesimals can occur frequently and have to be resolved with a very small time step. These two effects already make the problem very challenging with a small number of bodies.
Some of the first simulations of planetesimal dynamics in the inner solar system were able to integrate 100-200 bodies for $10^4-10^5$ yr \citep{ChambersWetherill1998,Aarseth+1993,Agnor+1999,Chambers2001}. The evolution of planetesimals and formation of planets was found to be a highly stochastic process and the solar system could not be reproduced well. A larger simulation containing $N\sim10^6$ planetesimals run for only 1000 yr by \cite{Richardson+2000} confirmed oligarchic growth, but was too short to study the entire planet formation process. The oligarchic growth was also studied in more detail over $4 \times 10^5$ yr with $10^4$ planetesimals by \cite{KokuboIda2002}.

The process of terrestrial planet formation over more than 100 Myr was studied by simulating the dynamics of 1000-2000 planetesimals by, e.g., \cite{Raymond+06}, \cite{Obrien+06} and \cite{Morishima+2010}, by including an analytic gas disk model into the integration. These simulations can also be used to estimate the delivery rate of water or other volatile elements  to the planets \citep{Elser+2012}. Since the full process of planet formation is stochastic, one cannot trust only one single simulation, but one has to study the statistics of many simulations with different initial conditions \citep{Kokubo+2006}. 
An overview of terrestrial planet formation with $N$-body simulations can be found in \cite{ChambersBook2011}.

With the discovery of more and more exoplanetary systems with more than one planet, it was natural to study their dynamics and stability in a similar way to the work done on the solar system. Test particles can be used to find stable islands between detected exoplanets, which can help to predict additional planets, while long term simulations can be used to constrain the orbital parameters of the planets by analyzing the stability of the system. Many exoplanetary systems have been studied, e.g., by \cite{MenouTabachnik2003,Asghari+2004,RaymondBarnes2005}, and the here presented code was already used in \cite{Elser+13} to study the stability of hypothetical super Earths in the habitable zones of exoplanetary systems.

\subsection{Technical Motivation}
Since $N$-body simulations can require a large amount of computing power, it makes sense to use the fastest computer systems available, in order to save computing time. A review about used hardware in the history of $N$-body simulations can be found in \cite{BedorfZwart2012}. Two highlights in the history of special purpose computers for $N$-body simulations are the ``digital orrery''  \citep{Applegate+1985}, and the family of  GRAPE (GRAvity PipE) computers \citep{Grape}.
The digital orrery was a special machine built to integrate the equations of motion of planetary systems similar to the solar system. It consisted of a ring of processors, each one computing the trajectories of one planet. This machine was used to perform the integration from \cite{SussmannWisdom1988}.
The GRAPE computers were able to compute the Newtonian force between two pairs of bodies directly in hardware. It was used as an accelerating device, sending the computed force between two particles to a central computer, on which the actual integration was performed. A short description of the different GRAPE types can be found in \cite{Grape}. 
Today's most efficient devices for $N$-body simulations are the GPUs. They consist of a large number of computing cores which can perform the same instructions on multiple threads (SIMT) in parallel. NVIDIA's GPUs can be programmed with the CUDA language (Compute Unified Device Architecture). Earlier methods to program GPUs are described in \cite{BedorfZwart2012}.

First results of GPU-based general type $N$-body simulations were published by \cite{PortegiesZwart+2007}, \cite{Belleman+2007} or \cite{HamadaIitaka2007}. A modern implementation of the gravitational force calculation on GPUs, optimized for $N$ $>$ 1024, are described by \cite{Gems3} or \cite{CudaHandbook}.
Codes using a Hermite integrator with block time steps for the general type $N$-body problem are given by the Sapporo library \citep{Sapporo}, the HiGPUs Code \citep{HiGPUs} or the NBODY6 Code \citep{NBODY6}.

A library supporting the parallel integration of small $N$-body Keplerian systems is given by SWARM-NG \citep{SwarmNG}.

\subsection{Contrast in Requirements with General $N$-body Simulations}
\begin{figure}
\epsscale{1.25}
\plotone{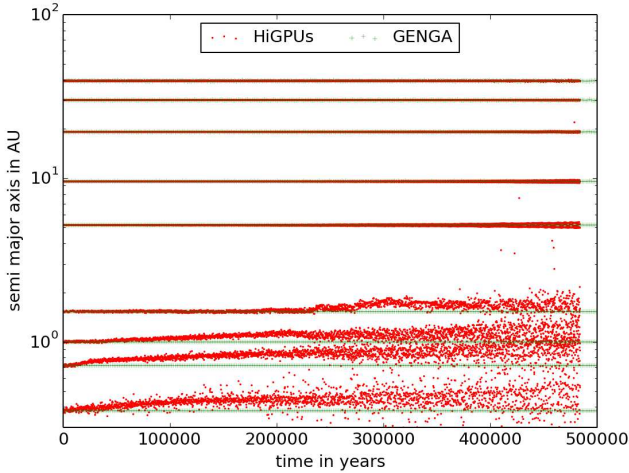}
\caption{Evolution of the semi-major axes of the solar systems planets, integrated with the two codes GENGA and HiGPUs. While GENGA is designed exactly for planetary system integration and conserves the energy of the system very well, HiGPUs solves the general type $N$-body problem and the relative error in the energy is 50,000 times larger than with GENGA. As a consequence the inner planets begin to drift away from the Sun. To perform the shown integrations, HiGPUs needed about 50 times more execution time than GENGA.}
\label{fig:HiGPUs} 
\end{figure}

The above listed codes are very efficient in solving the general type $N$-body problem, like in star clusters or cosmology simulations, where the individual bodies follow strongly non-Keplerian orbits. In planetary simulations by contrast, we can make the assumption that all bodies orbit a central mass following largely Keplerian arcs to lowest order with higher-order corrections resulting from mutual perturbations, where the solution of the Kepler problem can be computed analytically. Additionally, when a sufficient number of planetesimals are present within a system, close encounters may occur very frequently. The challenge is to conserve energy on secular timescales and yet treat close encounters properly, which is the focus of the present code. In order to demonstrate the importance of good energy conservation, we integrated the solar system using the HiGPUs code and compare the results with GENGA. In Figure\,\ref{fig:HiGPUs} is shown the evolution of the semi-major axes of the planets from the solar system. On the timescale of 500,000 yr, as shown in the Figure\,\ref{fig:HiGPUs}, the semi-major axes should remain almost constant, as reproduced by GENGA. In contrast the results of the HiGPUs code show a gradual but nonnegligible drift in the semi-major axis of the inner planets. Since the typical simulation for planet formation from planetesimals is typically integrated over 250 million yr, it makes it clear that a general $N$-body method should not be used for this problem.
The HiGPUs code uses a high-order Hermite integrator with block time steps and integrates the general $N$-body problem without the assumption that the gravitational force of the central mass is dominant most of the time. To integrate the first 10,000 yr, HiGPUs needs about eight times more execution time than GENGA. For the rest of the simulation it needs around 50 times more time than GENGA. However, this is not to say that the HiGPUs code is inefficient in its typical application domain. In fact it is one of the most efficient codes, solving the classical gravitational $N$-body problem by using CUDA together with OpenMP and MPI.

\subsection{Layout of the Paper}

The structure of this paper is as follows. In Section\,\ref{symplectic} we describe the theory behind the GENGA integrator and the used numerical methods, followed by a description of the GPU in Section\,\ref{GPU} with some considerations that need to be made when writing parallel code for these devices. In Section\,\ref{GENGA kernels} we give an overview of the different kernels and a detailed description of the implementation. In Section\,\ref{Results} we compare GENGA with Mercury and pkdgrav2 regarding the energy conservation, a planet formation test and the performance. In Section\,\ref{Limits} we show and explain the limitations of the current code.

\section{Theory, Formalism and Algorithms}

\label{symplectic}
The integration scheme of GENGA is based on the Mercury code \citep{Chambers99} and is a hybrid symplectic integrator  which can integrate close encounters between planetesimals with good long term energy conservation. Conservation of energy over a very large number of dynamical time steps is an important measure of the quality of an $N$-body integration. In the context of planetary system simulations, a drift in the energy results in an equivalent drift away from or toward the central mass: a qualitatively unphysical behavior in the numerical system.

\subsection{Background: The Second Order Hybrid Symplectic Integrator} 

An integrator which conserves the energy very well is the mixed variable symplectic integrator (MVS), described in \cite{WisdomHolman91}. It splits the gravitational Hamiltonian into a Keplerian and a perturbing part which can be solved independently. This Hamiltonian splitting is possible when using Jacobi coordinates. The Keplerian part can be solved analytically for each body separately, while the perturbing part involves calculating the accelerations between all pairs of bodies. The restriction of the MVS integrator is that the Keplerian part of the Hamiltonian must always be much larger than the perturbing part, which is fulfilled in most of the cases but not if some of the 
bodies are involved in a close encounter. Simply reducing the time step during a close encounter in the MVS integrator is not allowed since this would break the symplectic property of the numerical system and drifts (or jumps) in the energy would result. Another way to understand this is that for a symplectic integrator there exists an error Hamiltonian whose value depends on the chosen time step. In order to conserve energy, this error Hamiltonian must remain conserved and thus the time step must remain constant.

A solution to the close encounter problem is described in \cite{DuncanLevisonLee98} with the introduction of democratic coordinates, which consist of heliocentric positions and barycentric velocities. Using these coordinates the Hamiltonian is split up into three parts. In addition to the Keplerian part $H_A$ and the perturbing part $H_B$ one has also a ``Sun part'', $H_C$, which distributes the momenta of the central mass to all the other bodies. The advantage of these coordinates is that they do not depend on the order of the planets as is the case for the Jacobi coordinates. This permits individual parts of the Hamiltonian to be modified without affecting all the other bodies. For example, if a close encounter occurs then only the involved bodies need to be treated in a special way, by shifting those growing interaction terms from $H_B$ to $H_A$. 

\cite{DuncanLevisonLee98} use a hierarchical time stepping method to resolve the close encounters, by subdividing each time step into three smaller steps recursively for close interacting particles. This method decomposes the interaction potential into a series of matched functions, each applying to a different level of the time step hierarchy. A simpler method to integrate the close encounters is described by \cite{Chambers99} and implemented in the Mercury Code. A direct numerical integration (to machine precision) of the three or more body problem is used to handle $H_A$ during close encounters. While this method is less adaptive it is generally more efficient when close encounters are relatively infrequent. Chambers introduces a changeover function $K(r_{ij})$ which smoothly transfers the large parts from the perturbing part of the Hamiltonian to the Keplerian part. The new parts of the Hamiltonian are given as

\begin{eqnarray}
\label{Ha}
H_{A} = \sum_{i=1}^{N} \left( \frac{p_{i}^{2}}{2m_{i}}  - \frac{G m_{i} m_{0}}{r_{i0}} \right) \nonumber \\
- \sum_{i = 1}^{N} \sum_{j = i+1}^{N} \frac{G m_{i} m_{j}}{r_{ij}} [ 1 - K(r_{ij})]
\end{eqnarray}

\begin{equation}
\label{Hb}
H_{B} = -\sum_{i = 1}^{N}\sum_{j=i +1}^{N} \frac{G m_{i} m_{j} }{r_{ij}} K(r_{ij})
\end{equation}
\begin{equation}
\label{Hc}
 H_{C} = \frac{1}{2m_{0}}\left( \sum _{i =1} ^{N} \mathbf{p}_{i} \right) ^{2},
\end{equation}
where $H_A$, $H_B$ and $H_C$ are the Keplerian, perturbing and Sun part of the Hamiltonian in democratic coordinates. The index $0$ refers to the central mass.
Chambers defines the changeover function as

\begin{eqnarray}
 K(r_{ij}) = 
\begin{cases}
 0 &, y < 0 \\
 \frac{y^{2}}{2y^{2}-2y +1} &, 0 < y <1 \\
 1 &, y > 1
\end{cases}
\end{eqnarray}

with
\begin{equation}
y = \frac{r_{ij} - 0.1 r_{\mathtt{crit}}}{0.9 r_{\mathtt{rcrit}}}.
\end{equation}

If $K$ is equal to 1 then this formalism corresponds to the MVS integrator in democratic coordinates. The critical radius $r_{\mathtt{crit}}$ is the maximum of the critical radii $r_{\mathtt{crit,i}}$ and $r_{\mathtt{crit,j}}$ of the two bodies $i$ and $j$, with 
\begin{equation}
\label{rcrit}
r_{\mathtt{crit,i}} = \max(n_1 R_{H,i}, n_2 \tau v_i),
\end{equation}
where $ R_{H,i}$ is the Hill radius of body $i$ and $\tau$ is the time step. The two parameters are usually set to $n_1 = 3$ and $n_2=0.4$. The definition given by Equation\,(\ref{rcrit}) is slightly different from the definition used by Chambers. He uses the maximal velocity over all bodies instead of the velocity of the current body. We use the above definition in order to reduce the number of false positives in the close encounters detection. The velocity condition in the critical radius is needed to make sure that a minimum number of time steps are taken through the change-over function, such that the transition of the interaction terms from $H_B$ to $H_A$ proceeds smooth enough to bound the error in the energy.

With the Hamiltonian in the form given by Equations (\ref{Ha})-(\ref{Hc}), the second order solution of a phase space vector $z = (\mathbf{q}, \mathbf{p})$ is:
\begin{equation}
\label{z1}
z(\tau) = e^{\frac{\tau}{2} B} e^{\frac{\tau}{2} C} e^{\tau A} e^{\frac{\tau}{2} C} e^{\frac{\tau}{2} B} z(0).
\end{equation}
For example the operator $A$ can be computed with the formula:
\begin{equation}
\label{O}
 \frac{dz}{dt} = \sum_{i = 1}^{3N} \left( \frac{\partial \mathbf{q}}{\partial x_{i}} \frac{\partial H_A}{\partial p_{i}}  - \frac{\partial \mathbf{p}}{\partial p_{i}} \frac{\partial H_A}{\partial x_{i}}\right) = Az.
\end{equation}

\subsection{Algorithms}

The Formula\,(\ref{z1}) describes an algorithm to evolve the bodies for one time step $\tau$. The first step is to calculate the accelerations between all the bodies, and to apply a velocity kick operation for half of the time step. The second step is to compute the total momentum of the system and distribute it to the bodies to adjust the system such that the velocities of the bodies are barycentric. The third step is to move the bodies for one time step along Keplerian orbits around the central mass, where the Keplerian orbit can be computed analytically. Then the steps two and one are repeated.

To move bodies along a Keplerian arc, we use Gauss' $f$ and $g$ function method as described in \cite{Danby88}. This involves solving Kepler's equation in differential form to obtain the functions $f$ and $g$ for a given time interval $\tau$ from which the new position and velocity of the body are given by
\begin{equation}
\label{fg}
{\bf x}_\tau = f_\tau {\bf x}_0 + g_\tau {\bf v}_0
\hspace{0.1in}
{\bf v}_\tau = \dot{f}_\tau {\bf x}_0 + \dot{g}_\tau {\bf v}_0 .
\end{equation}
The FG method is computed in democratic coordinates, which means that the reduced mass here is given by $\mu = G M_{\odot}$ and is not $\mu = G (M_{\odot} + M_i)$, as is usually used.

When a body is in a close encounter, then the part $H_A$ cannot be computed analytically and the affected bodies have to be integrated with a direct $N$-body integrator. For the direct $N$-body integration we use a Bulirsch-Stoer method, as recommended by \cite{Chambers99}. Compared to a Hermite Predictor Corrector Scheme \citep{MakinoAarseth92} and \citep{NitadoriMakino08}, a higher order Runge Kutta Fehlberg method \citep{Hairer} or a Lie series integrator \citep{HanslmeierDvorak84}, the Bulirsch-Stoer integrator shows the best performance and accuracy for our problem.

Additional steps in the algorithm are the search for close encounters and the grouping of independent close encounter pairs. As described in \cite{Chambers99}  polynomial interpolation can be used to find all close encounter pairs in a time step, by using a cubic Hermite polynomial of the form
\begin{eqnarray}
 P(t) = P(0) (1+2t)(1-t)^2 + P(1) t^2(3-2t) +  \nonumber \\
 \dot{P}(0) t(1-t)^2 \tau + \dot{P}(1)t^2(t-1)\tau,
\end{eqnarray}
where $P(0), \dot{P}(0), P(1)$ and $\dot{P}(1)$ are the square of the difference of the positions and velocities between two bodies at the beginning and the end of a time step. The parameter $t$ has a value between zero and one. To find the minimal distance of the two bodies within a time step, one sets
\begin{equation}
\label{teqn}
 \frac{dP(t)}{dt} = at^2 + bt + c \ \dot{=} \ 0,
\end{equation}
for
\[
 a = 6(P(0) - P(1)) + 3 \tau(\dot{P}(0) - \dot{P}(1)),
\]
\[
 b = 6(P(1) - P(0)) - 2 \tau(2\dot{P}(0) + \dot{P}(1))
\]
and
\[
 c = \dot{P}(0) \tau.
\]
Solving Equation\,(\ref{teqn}) for $t$ gives the square of the minimal distance between the two bodies as
\begin{eqnarray}
\label{polydelta}
 \Delta = (1-t)^2(1 + 2t)P(0) + t^2(3 - 2t)P(1) + \nonumber \\
 t (1-t)^2 \tau \dot{P}(0) - t^2 (1-t) \tau \dot{P}(1).
\end{eqnarray}

Checking all $N(N-1)/2$ possible pairs of bodies for close encounters using this polynomial fitting method would be too time consuming and wasteful 
since it is possible to first cull out the majority of pairs which cannot have an encounter from much simpler considerations.  
This culling of the number of potential close encounter pairs in the job of the prechecker which is performed during the kick operation. As discussed later in the section describing the Bulirsch--Stoer direct integration, during such close encounters actual collisions between the bodies are possible.
In GENGA the prechecking is done by increasing the critical radius by a factor of three and comparing this enhanced critical radius to the separation of the bodies at the start of the time step. This approach is very simple and efficient since we already calculate all pairwise distances within the kick kernel. The usage of the critical radius as a precheck separation limit is possible because it depends on the velocity and the time step (Equation\,(\ref{rcrit})). It sets a maximum distance which a body can move in the next time step. Other prechecking techniques would be possible for example, comparing the perihelion and aphelion of the bodies and/or including the phase of the orbit. Our simple prechecker usually reports around 10 times more candidates than confirmed close encounters which is adequate to a number of bodies up to 2048.

If no close encounters occur, then the opening kick operation of the next time step is identical to the closing kick operation of the current time step, and they can be combined. The very first opening kick operation of the simulation has to be computed separately.
The complete algorithm for one time step $\tau$ is the following, in the case of no close encounter candidates:
\begin{itemize}
 \item compute critical radii
 \item do opening kick for $\frac{\tau}{2}$ by using the known accelerations from the last closing kick
 \item do Sun kick for $\frac{\tau}{2}$
 \item do Keplerian drift for $\tau$
 \item do Sun kick for $\frac{\tau}{2}$
 \item do closing kick using newly calculated accelerations for $\frac{\tau}{2}$ and do the precheck search.
 
\end{itemize}

In the case of close encounter candidates, the algorithm looks as follows:
\begin{itemize}
 \item compute critical radii
 \item do opening kick for $\frac{\tau}{2}$ by updating only the accelerations of the close encounter candidate pairs
 \item do Sun kick for $\frac{\tau}{2}$
 \item do Keplerian drift for $\tau$
 \item do close encounter search
 \item if close encounters:
 \begin{itemize}
 \item do grouping
 \item do direct integration for $\tau$
 \item do collisions
 \end{itemize}
 \item do Sun kick for $\frac{\tau}{2}$
 \item do closing kick using newly calculated accelerations for $\frac{\tau}{2}$ and do the precheck search.
 
\end{itemize}

\subsection{Generalization to Higher Orders}
From the second order solution given in Equation\,(\ref{z1}), higher order integrator schemes can be constructed as described in \cite{Yoshida90}. In the higher order symplectic schemes, each time step is split up into more sub-steps, where some of them can also be backward in time. In the higher order symplectic hybrid integrator one cannot simply use the second order close encounter search over one full time step, because the opening and closing kick operations would not be synchronous with the desired coordinates for doing the polynomial fitting described in Equation\,(\ref{polydelta}). For practical reasons we apply the full scheme, including the close encounter search and the Bulirsch-Stoer integration, but not the prechecker to each of the substeps.
To perform only one close encounter search over the full time step would require additional drift and kick operations and also a higher order polynomial fitting.
GENGA supports fourth and sixth order symplectic schemes, where it uses the solutions A from \cite{Yoshida90} for the sixth order scheme.

\subsection{The GPU and CUDA}
\label{GPU}

A GPU consists of a large number of cores which can perform the same instructions on multiple threads (SIMT) in a very efficient way. These parallel executed code sequences are called threads. The threads are grouped together logically into three dimensional thread blocks and the thread blocks themselves are grouped in a two dimensional grid. This structure is a purely logical organization and is not related directly to how hardware is organized on the physical GPU device. All the threads have their own local memory and registers, and in addition every thread block has a very fast shared memory. Each thread block can contain up to 1024 threads, depending on the GPU generation. The data on the GPU is stored in a global memory, to which all threads have access. Reading from global memory is slow and should be reduced to a minimum. The global memory of a modern high end GPU has a size of up to 6 GB.  

In addition to the logical structures of the grid and thread blocks, there exists another important unit given from the hardware: the warp. The warp is a unit of 32 threads and is the smallest parallel execution unit following the SIMT concept. All 32 threads in a warp must perform the same instructions and are always executed synchronously. To achieve good performance, branch divergences within a warp should be avoided, and a block size should always be a multiple of the warp size.

An important bottleneck in GPU computing is the data transfer between the CPU and GPU. If one uses the GPU as an accelerator which handles only some computationally intensive parts of the problem while the rest is left to the CPU, or if the problem is too large to fit into the global memory of one GPU, then one has to hide the memory transfer behind other operations. When the problem is small enough to fit completely into the GPU memory, one can avoid the bottleneck by performing the simulation entirely on the GPU. This means that all parts of the code need to be parallelized according to the CUDA programming model (NVIDIA CUDA C Programming Guide\footnote{http://docs.nvidia.com/cuda/cuda-c-programming-guide}), which may not be the most efficient one for some routines. In GENGA this is probably the case for the group finding algorithm, but since this part is only secondary with respect to the kick part, and since we want to focus on a number of bodies not higher than 2048, we decided to implement GENGA as an entirely GPU code. In order to control and synchronize the operations, only a very small amount of data needs to be transferred between the CPU and the GPU. To not affect the performance, output files should not be written too frequently.

Compared to a modern CPU, the GPU has a lower clock rate, and a function call on the GPU, called a kernel launch, will cause much more overhead time than a CPU function call. This means that the GPU needs a minimum amount of parallel work to be able to hide this overhead time. Simulating a system with only a small number of bodies will never be as fast on a GPU as on a CPU. But if the number of bodies is large or if we simulate many small systems in parallel, then the GPU can become very efficient. To reduce the amount of kernel launches, one could try to write the full code as only one kernel, but since different parts of the code will need a different parallelization structure, this way would not be very efficient.

During a simulation the number of bodies will decrease with time due to collisions and ejections, which means that the computational kernels must cover a large range in the numbers of bodies. For some kernels it's not possible to write only one version which covers the full range, due to limited shared memory or a limited number of threads per block. For some other kernels it's possible to use the same code but with a different amount of shared memory. If the number of bodies goes below a certain limit, then a different set of kernels is launched. This adaptive parallelism maximizes the effectiveness of the GPU over a range in $N$ from 16 to 2048 bodies.

\section{Structure of the GENGA Code}
\label{GENGA kernels}
GENGA supports three simulation modes. The main mode integrates a planetary system with up to 2048 massive bodies orbiting around a central mass. The test particles mode integrates up to one million massless test particles in the presence of maximally 32 massive bodies. The multi-simulation mode integrates up to 100,000 independent planetary systems each with no more than 16 bodies.

\subsection{Overview of the Different Kernels}

The different operations of the integration scheme are split up in different kernels or functions. Here we give a short description of the most important ones. A detailed description of the kernels is given in section\,\ref{details}.
\begin{itemize}
 \item Rcrit: it calculates the critical radius of all bodies (Equation\,(\ref{rcrit})).
 \item FG: it drifts the bodies along Keplerian arcs using the Gauss' $f$ and $g$ function method (Equation\,(\ref{fg})).
 \item HC: it applies the Sun kick operation, by calculating the total momentum of the system (Equation\,(\ref{Hc})).
 \item Kick: it applies the kick operation by calculating the accelerations between all bodies. It includes also the prechecker for the close encounter detection. The kick kernel is not used in the first kick operation of a time step.
 \item KickA: it is used in the first kick operation of a time step, in the case of close encounters. It reuses some accelerations computed in the last kick kernel of the previous time step.
 \item KickB: it is used in the first kick operation of a time step, in the case of no close encounters. It reuses all accelerations computed in the last kick kernel of the previous time step.
 \item Encounter: it calculates the minimal distance of all pairs of bodies (Equation\,(\ref{polydelta})) marked by the prechecker in the kick kernel. The Encounter kernel creates a list of all close encounter pairs.
 \item Group: it finds indirect close encounter pairs and separates independent close encounter groups.
 \item Fusion: it is used to merge together large close encounter group lists.
 \item BS: The Bulirsch-Stoer direct integration of close encounter groups.
 \item Other less apparent operations which lead to significant overheads in the parallel implementation are Sync, which synchronizes the GPU with the CPU  and Copy, which transfers some information about the number of threads from the GPU to the CPU. These are shown in Figure\,\ref{fig:Timing}.
\end{itemize}

In order to treat a variable number of bodies and to support the different computing modes of GENGA, different versions of the listed kernels are needed. In Table\,\ref{tab:kernels} is shown the number of different versions for the main kernels.

\begin{table}
\begin{center}
\begin{tabular}{c|c}
Kernel & Versions \\
\hline
Rcrit & 3\\
FG & 3\\
HC & 5\\
Kick & 7 \\
KickA & 2\\
KickB & 2\\
Encounter & 3\\
Group & 5\\
Fusion & 5\\
Energy & 4\\
BS & 6\\

\end{tabular}
\end{center}
\caption{An overview of the different kernels with the number of implemented versions.}
\label{tab:kernels}
\end{table}

\subsection{GPU Implementation Details}
\label{details}

\subsubsection{The FG Kernel}
\label{FG}
The parallelizations of the FG kernel is very simple because there are no dependencies between the bodies. One can simply use one thread per body and use shared memory to speed up the operations, but some attention is needed for an accurate implementation.
During the calculation of the FG method, one needs to apply the sine and cosine functions, which can be computed simultaneously with the $sincos()$ function. In single precision CUDA supports the very fast intrinsic function $\_\_sincosf()$, but the result is not an IEEE standard and not accurate enough for achieving long term energy conservation. Details about the intrinsic functions can be found in the NVIDIA CUDA C Programming Guide\footnote{http://docs.nvidia.com/cuda/cuda-c-programming-guide}.
Another very important operation is the calculation of the inverse distance $r_{ij}^{-1}$, which could be calculated using the fast $rsqrt()$ function. However, this function is also not accurate enough and it can cause growing errors in the integration. We therefore always use the usual $sqrt()$ function combined with an additional division operation. The FG kernel needs a large number of registers and the CUDA compiler tries to optimize the code by combining some operations. On some compiler versions these optimizations can also lead to computing errors. To avoid these errors it is necessary to specify some of the longer expressions as being of volatile type, at the expense of performance.

The FG method needs a few iterations to converge. In most of the cases three or four iterations are enough to converge to machine precision. One could imagine performing the first one or two iterations in single precision only using the intrinsic functions for sine and cosine, and then continue the last iterations in double precision, but timing experiments have shown that it is faster to perform all iterations directly in double precision, and using a simpler algorithm. If a body has a very high eccentricity, it can happen that the FG method does not converge at all. It can also happen that the eccentricity is larger than one, indicating an unbound orbit, for which the conventional FG method breaks down (there are methods using universal variables that resolve this issue \citep{Danby88}). In these relatively rare cases we fall back on our Bulirsch-Stoer method to integrate the two-body problem. We describe the application of this method to close encounter orbits later in this section.
Since not all the threads need the same number of iterations to converge, the FG method can cause branch divergences. To minimize these branch divergences we use only 32 threads per thread block. 

\subsubsection{The HC Kernel}
\label{HC}

The main operation in the HC kernel is to calculate the total momentum of the system and to distribute it to all the bodies. A very efficient way of performing the summation over all bodies is to use a reduction formula as described by \cite{Harris07}. The summation can be done very rapidly in $log_{2}(N)$ steps using shared memory. Since the maximum number of bodies is only about four times higher than the maximum number of threads per block, we want to perform the full kernel in only one single block, with as many threads as possible. To cover the full range of bodies we have to include a serial loop in the kernel. If the maximum number of bodies is much larger than the maximum number of threads per block, it would be better to use more than one block, but then the synchronization would be more complicated.
 
By passing the kernel a template argument with the current number of bodies, the compiler can reduce the formula to the right number of steps. The last six steps of the reduction formula are performed all in the same warp, which is a group of 32 threads that must all perform the same instructions on different data within the hardware (SIMT). Therefore the last six steps do not need any synchronization on current GPU generations, but on Fermi and Kepler type cards, we have to use a volatile type for the variables to get the correct result.
Not using a volatile type in the last reduction steps will cause an error on the new cards because the compiler will try to optimize the code too much and will not update all the intermediate results.
 The result of the reduction formula is stored in thread number zero and is distributed to all the other bodies by using a broadcast.

\subsubsection{The Kick Kernel}
\label{kick}
The main work in the kick operator is to compute the accelerations between all pairs of bodies. In our code we use a direct summation technique to compute the force acting on all bodies. We could imagine using some more complex techniques with a lower order of operations like a fast multipole tree code or a spherical harmonic expansion code (both scale as $O(N)$), but in a range of up to a few thousand bodies these techniques would not be faster on the GPU than direct summation ($O(N^2)$).

A description of gravitational force calculation, optimized for a number of bodies larger than 2048 is given in \cite{Gems3} or \cite{CudaHandbook}. These implementations split the $N^2$ interactions into small tiles which can be stored in shared memory. Only one dimension of the $N^2$ interactions are performed in parallel, the other dimension is performed sequentially. The descriptions in \cite{Gems3} provides additionally a version optimized for a smaller number of bodies which computes also some parts of the second dimension in parallel. For a number of bodies smaller than 512 this method still does not provide enough parallel work for the GPU to work efficiently. For that reason we implemented a different version of the kick operation using more parallel work. We use a reduction formula as in Section\,\ref{HC} to perform the summation $\mathbf{a}_{i} = \sum_{j} \mathbf{a}_{ij}$ of all the interactions.
For a small number of bodies we therefore use $N$ blocks each with $N_b$ threads, where $N_b$ is the next power of two larger than $N$. Timing experiments have shown that for a larger number of bodies it is faster to compute more than one body within each thread block. Using too many thread blocks increases the kernel overhead, using too few causes bank conflicts and uses too much shared memory. In Table \ref{tab:Nb} is shown how many bodies that every thread block should compute accelerations to get the best performance. These results are based on timing experiments on a GTX 590 card, on newer cards this result can be different.
If the number of particles exceeds the maximum number of threads per block, then the computation of the accelerations $\mathbf{a}_{ij}$ are embedded in a for loop with a step size equal to the block size. Each iteration in this loop computes $\mathbf{a}_{ij}$ for a consecutive bunch of bodies $j$, while the index $i$ is still given by the block index.
Another speed up in the range of bodies in $ 512 \leq N \leq 2048$ can be achieved by splitting the number of threads per block in two halves. One half computes the acceleration $\mathbf{a}_{ij}$ and $\mathbf{a}_{(i + N/2) j}$, and the other half  $\mathbf{a}_{(i+ N/4)j}$ and $\mathbf{a}_{(i + 3N/4) j}$, for $ 0 \leq i \leq N/4$.

With our implementation it was easy to include the  needed second acceleration array, described in the next section, and the close encounter count function. But we have to admit that for more that 512 bodies an implementation similar to \cite{Gems3} would lead to a performance improvement of a few percent. This change is planned for future versions of the code.

As described in Section\,\ref{FG}, we do not use the $rsqrt()$ function to avoid computation errors. Even though these errors in $rsqrt()$ are only very small, if they are biased in any way, they could cause a spurious numerical drift in quantities and hence could qualitatively change the final result of a simulation after a very large number of time steps. Since we cannot completely rule out such a possibility at the moment, we chose to take the more conservative IEEE-754 compliant $sqrt()$ and division option at a slight performance penalty.

\begin{table}
\begin{center}
\begin{tabular}{c|c}
$N_b$ & $n_i$\\
\hline
32 & 1\\
64 & 1\\
128 & 2\\
256 & 4\\
512 & 4\\
1024 & 4\\
2048 & 4 \\
\end{tabular}
\end{center}
\caption{The structure of the kick kernel. Listed are the number of bodies, which are computed in one thread block $n_i$, as a function of the total number of threads per block $N_b$.}
\label{tab:Nb}
\end{table}

\paragraph{Combining the Kick Kernels}

In the second order integrator, each time step contains two kick operations, the opening kick at the beginning and the closing kick at the end of a time step. In higher order integrators, there are more kick operations in between. The closing kick operation and the opening kick operation of the next time step differ only in the updating of the critical radius of each body. Most of the bodies will not be in a close encounter, neither in the current time step, nor in the next one, which means that the accelerations between all such pairs remain the same and need not be updated. Only pairs of bodies involved in a close encounter or near to a close encounter can have differing accelerations from one step to the next, due to a change in the change-over function. These pairs will be detected by the prechecker during the kick operation.
The acceleration between all pairs of bodies which are not reported by the prechecker, can therefore be reused in the next opening kick.
If there are no close encounter candidates, then in the next time step the kickB kernel is launched, which kicks the bodies with the already known acceleration. This kernel is very simple and fast.

If there are some close encounter candidates, then the kickA kernel is launched, which reuses the accelerations of the bodies not marked from the prechecker, and calculates only the missing accelerations of the close encounter candidates.
In order to decide which kernel to launch, the CPU has to know the number of close encounter candidates. This can either be done by using
mapped memory, followed by a synchronization function on the CPU. But in this case it is faster to use a memory copy function to transfer the value from the GPU to the CPU without calling a synchronization function.

\subsubsection{The Encounter Kernel}

To find the real close encounter pairs from the candidate list, we use the same cubic Hermite spline interpolation as described in \cite{Chambers99}. The encounter kernel uses one thread per candidate pair to find the real close encounters. Since the encounter kernel can cause branch divergences like the FG kernel, depending if the candidates are confirmed or not, we use only 32 threads per block. The CPU already knows the number of close encounter candidates for launching the kickA or kickB kernel, and can use here the same number to determine the number of thread blocks. The confirmed close encounter pairs are written into an array, and similar to the prechecker we use an $atomicAdd$ function to calculate the total number of close encounter pairs.

\subsubsection{Grouping the Close Encounter Pairs}
\begin{figure}
\plotone{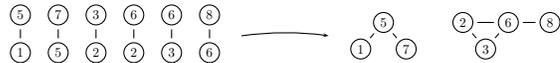}
\caption{Starting from a list of close encounter pairs, drawn at the left hand side, the parallel searching algorithm finds indirect close encounter pairs (e.g., the pairs 1 and 7) and separates the list into individual close encounter groups which can be integrated in parallel.}
\label{fig:group} 
\end{figure}

In the encounter kernel we created a list of all close encounter pairs, which will be integrated with the Bulirsch-Stoer method. If all the bodies would be in a close encounter with only one other body, then we could integrate all these pairs independently and fully in parallel. But it can happen that some bodies are in a close encounter with more than one body, and create indirect close encounter pairs which should be concatenated into bigger groups as illustrated in Figure\,\ref{fig:group}. By using a large number of bodies and a big critical radius, these groups can reach very large sizes as shown later in Figure\,\ref{fig:cluster}.

The group kernel finds all indirect close encounter pairs using a parallel searching algorithm using shared memory. This is the same as the standard method \citep[Sections~4.6 and 5.81]{HorowitzSahni} for determining equivalence classes from a set of equivalence relations. The only difficulty is that these classes (or encounter groups in our case) must be constructed in parallel. Consistency across all the parallel threads of a block is achieved by using an atomic minimum operation (a CUDA primitive function) to update the equivalence classes. Furthermore one needs to iterate these updates until no more changes occur in order to construct globally consistent equivalence classes. Details to this algorithm can be found in Appendix\,\ref{appendix}.

\subsubsection{Bulirsch-Stoer Integration}
The close encounter groups are integrated in a way similar to Mercury with the Bulirsch-Stoer method, but in GENGA we want to integrate the different close encounter groups in parallel. The computational flow can vary a lot between different close encounter groups, depending on the size and the nearest distance between the bodies. Therefore, we want to use one block for each group, and use as many threads as reasonable. We divide the close encounter groups in different size classes, where the sizes are set by powers of two. Each size class is launched with a different stream and with a fixed number of threads, given in Table\,\ref{tab:BS}.

In the Bulirsch-Stoer kernel the accelerations between all pairs of bodies $\mathbf{a}_{ij}$ are computed. To be able to use as many threads as possible in parallel we define the indexes $i$ and $j$ as follows through the thread index $\mathtt{th}_i$ and a parameter $p$ which sets the amount of parallelism:
\[i = \mathtt{th}_i / p\]
\[j = \mathtt{th}_i \% p,\]
where we used in the first line an integer division and in the second line the modulo operator. 
If the class size is bigger than the parameter $p$, then we loop around the remaining pairs by setting $j = j + k* p$. To sum up all $\mathbf{a}_{ij}$ terms we use again a parallel reduction formula.

\begin{table}
\begin{center}
\begin{tabular}{c|c|c|c|c}
Group & Kernel& Stream& Threads& $p$\\
Size  & Name &	 & per Block &  \\
\hline
2 & BSB & 0 & 4 & 2\\
3-4 & BSB & 1 & 16 & 4\\
5-8 & BSB & 2 & 64 & 8\\
9-16 & BSB & 3 & 256 & 16\\
17-32 & BSB & 4 & 256 & 8\\
33-64 & BSB64 & 5 & 256 & 4\\
65-128 & BSB128 & 6 & 256 & 2\\
\end{tabular}
\end{center}
\caption{The parameters for parallelized Bulirsch-Stoer integration for different sizes of close encounter groups. The parameter $p$ sets the amount of parallelization in the kernel.}
\label{tab:BS}
\end{table}

If there are groups containing more than 128 bodies, the described method gets inefficient and would also use too much shared memory. In this case it's faster to split the Bulirsch-Stoer method into different kernels which perform separately the accelerations, error estimations and acceptances. These parts are then controlled by the CPU, which creates more kernel overhead time but can also use more threads for a better parallelization. 

Performing the Bulirsch-Stoer integration in democratic coordinates means that the position of the central mass is constant during one time step.

\subsubsection{Collisions}
During a close encounter it can happen that the separation between two bodies $r_{ij}$ gets smaller than the sum of their physical radii $R_i + R_j$, which means that the bodies collide. In the simplest model the two bodies collide as perfectly inelastic bodies, by forming one bigger body. Linear momentum is conserved during the collision but energy is not, since a part from the kinetic energy and the potential self-energy is transferred into an internal energy $U$. To be able to check the energy conservation during a full  simulation, we compute the internal energy from each collision as
\[ U =  \frac{1}{2} \frac{m_{i}m_{j}}{m_{i} + m_{j}} v_{ij}^{2} - G \frac{m_{i}m_{j}}{r_{ij}}. \]

The spin of the new body is calculated from angular momentum conservation. The index of the new body is the index of the more massive body, and if both bodies have an equal mass, then the smaller index is used. Technically we transform the body $i$ into the new body, and body $j$ into a massless ghost particle which will be removed later.

Each collision is reported by writing the positions, velocities and the spins at the last time step before the collision happens into a collision file. Since writing to a file is not possible from a kernel, the information is written into a buffer and then copied back to the CPU after the kernel is terminated.

To find collisions between the Bulirsch-Stoer time steps, we use the same code as in the close encounter detection. 

\subsubsection{Ejections}
A body is treated as ejected if the distance to the central mass exceeds the limit $Rcut$, or if it is too close to the central mass, specified by the limit $RcutSun$. In both cases the coordinates of the body are reported in an ejection file and the body is removed from the simulation.

\subsection{Test Particle Mode}
\label{TestParticles}
Test particles are massless particles with a physical radius greater than zero. The orbits of the test particles are perturbed by the planetesimals, but not by other test particles. When a test particle collides with a planetesimal, it writes a report and is removed from the simulation, it has no effect on the planetesimal. Since the test particles do not interact with each other, the computation time can be reduced significantly.

Many kernels of the test particle mode are very similar to the ones for massive bodies, like the FG, Rcrit, kickB and the encounter kernel. In the HC kernel the summation over the total momentum runs only over the massive bodies, but the result is then applied to the test particles as well. The biggest difference to the massive body kernels is in the kick kernel. This is designed to integrate a large number of test particles, interacting only with 32 massive bodies at maximum. The easiest way to parallelize this problem is to use one thread for each test particle and loop over the massive bodies. The massive bodies are integrated with the same kick kernel as in Section\,\ref{kick}. 

The easiest way to parallelize this problem is to use one thread
for each test particle and loop over the massive bodies.

The group finding scheme needs to be split up into three different kernels, due to synchronizations between the thread blocks. It can happen that many test particles are in a close encounter with the same planetesimal, but since the test particles do not affect other particles, we can integrate all of these close encounters independently as different groups containing only one test particle and the planetesimal. Only the test particle gets updated by the Bulirsch-Stoer integration, the planetesimal keeps the coordinates from the FG kernel. Close encounters between planetesimals are treated separately.

\subsection{Multi-Simulation Mode}
\label{MultiSimulation}
The multi-simulation mode can integrate a large number of independent systems in parallel, where each of these systems can have an individual number of bodies, but at maximum 16. The difficult part of running small independent systems in parallel is how to distribute them along the thread blocks. 

If all systems would have exactly 16 bodies, it would be very simple and we could just launch one block with 16 threads for each of the systems. However, most applications would probably consist of systems with only three or four bodies. To run all of them with 16 threads would be a waste of resources. In this section we describe a better solution on how to parallelize this problem.

\subsubsection{Organization and Memory Allocation}

Each simulation has its own directory, containing the initial conditions, the parameter file and all the output files. Not all parameters of the parameter file can be set individually for the different simulations, only the output name, the central mass, the $n_1$ and $n_2$ parameters, the input file name, the number of bodies and the minimal number of bodies.
All the other parameters are copied from the first simulation. The reason why these remaining parameters cannot be chosen individually, is because these will need more synchronizations and a larger memory transfer between the CPU and the GPU, which will slow down the code significantly. In order not to negatively impact the performance one should not write outputs too frequently.

The coordinates of the different simulations are all stored consecutively in the same array. Since the simulations can have an individual number of bodies, it is not trivial which bodies belong to which simulations. For this reason we create a new array, which contains the starting points of the different simulations. To be able to find the right parameters for each simulation, we need to add a simulation index, which can easily be included in the body index array by modifying the index internally as $i' = i + 100 \cdot si$, where $i$ is the body index and $si$ the simulation index. In the multi simulation mode, the body index cannot be greater than 100.

\subsubsection{Simple Kernels}
Some of the kernels have no dependency between the bodies, and are therefore very easy to parallelize by using just one thread per body. 
In this way we can perform the drift kernel, the Rcrit kernel, the kickB kernel and the encounter kernel.

\subsubsection{The HC Kernel}
In the HC kernel, we have to sum up the momenta for each system and to perform the kick operation on each of the bodies. The difficulty is that the number of bodies of each simulations is mostly smaller than a warp size and all systems can have an individual number of bodies. It would not be efficient to launch a different thread block for all simulations. Therefore we have to calculate more than one system within one block. To solve this issue, the summation can be calculated with the same reduction code in shared memory as usual, but here the summed momenta must be multiplied with zero when the two bodies do not belong to the same simulation. Once the sum is calculated, the result has to be distributed along the bodies with a reduce-scatter operation.

At the boundary of the blocks, we need to take a closer look. It can be that a system is split up into different blocks and therefore the total momenta cannot be calculated directly. For this reason we insert some ghost threads on both sides of the thread block to make sure that the threads near to the boundaries are computed correctly.

\subsubsection{The Kick Kernel}
Similar to the HC kernel we have to compute more than one simulation within a thread block and even within a warp. Here we cannot use a reduction code, because each body has an individual acceleration. We use again one thread per body, but loop around all the interaction partners of the current simulation. Since all threads within the same warp must perform the same instructions (SIMT), and simulations can have a different number of bodies, sometimes false interactions are computed, which are then taken out of the result.
Using this method, all threads can perform the same operations. Like in the HC kernel we include ghost threads at the boundaries to calculate split systems correctly.

\subsubsection{The Group Kernel and Bulirsch-Stoer Integration}

To compute the group index of each close encounter pair, we can use the same algorithm as before, but to create the lists containing the group sizes, a second kernel has to be launched because this is a different parallization problem and a different number of thread blocks is needed.

The Bulirsch-Stoer integration can as well be performed in the same way as before, because the members of the close encounter groups belong all to the same simulation. The only difference is the fact that different groups can have a different central mass, and that collisions are reported in different files. 

\subsubsection{The Remove and Stop Kernel}

A typical application of the multi-simulation mode might be to simulate many instances of a planetary system with a specific number of known planets and some hypothetical bodies in addition. In this situation, the result is only relevant if all of the known planets are still part of the simulation. As soon as one of the know planets gets ejected or if it collides, then the future orbits might not be interesting, and it makes sense to stop the affected simulation. For this reason we include the $Nmin$ parameter which sets a minimal number of bodies to the simulations.

When a body gets removed from a simulation due to a collision or an ejection, then the body gets deleted and the memory gets compacted by putting the last body of the affected simulation to the deleted position. The number of bodies of the affected simulation is reduced by one, but the starting points of the simulation in the memory and the total number of bodies still remains the same. In this way the data of the different simulations is still well separated in memory while the execution time can be reduced consecutively.

Only if the number of bodies in a simulation becomes smaller than the minimal number of bodies specified in the parameter file, will the simulation be stopped by deleting all the bodies of the affected simulation. The starting points in the memory then need to be updated and the total number of bodies is reduced. For simplicity  the recalculation of the starting points is done on the CPU rather than on the GPU. Performing this operation on the CPU makes it easier to reorganize the memory while the overall performance is not affected because this operation is called only very rarely. Alternatively a parallel implementation on the GPU would be possible by using a scan operation.

\section{Results}
\label{Results}

In this section we first quantify the energy conservation of GENGA, followed by a planet formation simulation in comparison to two other codes. Finally we analyze the performance of all parts of GENGA.

\subsection{Energy Conservation}
\label{Energy}
\begin{figure*}
\plotone{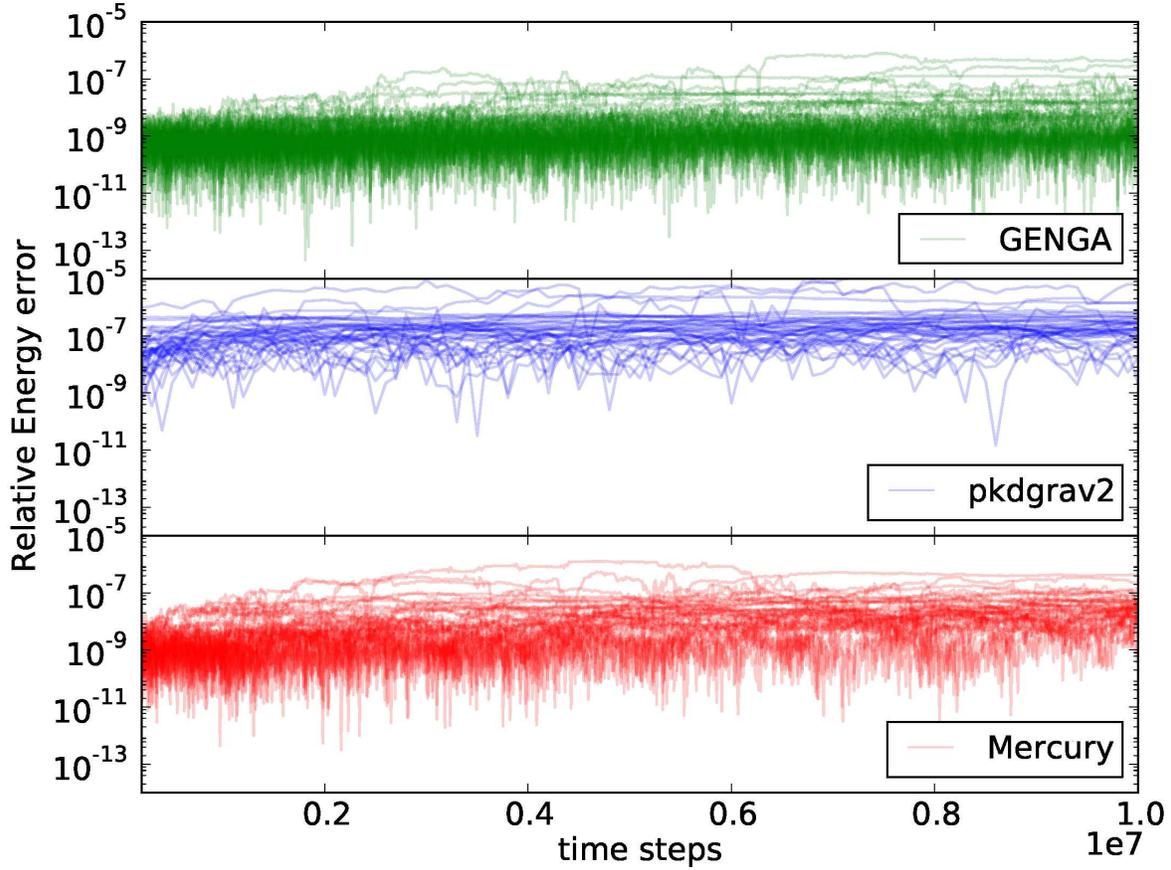}
\caption{Comparison of the relative energy error for the three codes, GENGA, pkdgrav2 and Mercury, for a set of 40 simulations with 32 bodies each, with a total mass of 5 Earth masses, distributed between 0.5 and 4 AU. The close encounter parameters used are $n_1 =3$, $n_2=0.4$ and the time step $\tau$ is set to 6 days. Over all 40 simulations the energy conservation of GENGA is somewhat better than Mercury, although the drift in the energy of the worst outliers is about the same for the two codes. The energy conservation of pkdgrav2 is less good, because it uses an approximated value of the forces between the bodies.}
\label{fig:MercuryEnergyP32} 
\end{figure*}

The quality of energy conservation of the hybrid symplectic integrator depends strongly on the initial conditions. If there are often transitions through the critical radius, defined by the Hill radius and the velocity, then the energy is not perfectly conserved. In Figure\,\ref{fig:MercuryEnergyP32} is shown the relative energy error $ E_{\mathtt{rel}} = |\frac{E_t - E_0}{E_0}|$ for a set of 40 simulations with 32 bodies each. The initial conditions of all simulations are drawn from the same distribution function with different random numbers. The total planetary mass of all the systems are 5 Earth masses, distributed between 0.5 and 4 AU. 
Most of the simulations show very good energy conservation over a long time scale, but for some of them the energy begins to drift away after two million time steps.
The main reason for the energy errors is the fact that the transition between the two integrators is not symmetric in time, the relative speed and the angle between the orbits of the two close encounter bodies are not equal in the approaching and receding transition (the gradual switch to the Bulirsch-Stoer integration and gradual switch back to the MVS integration). Basically, if the second time derivative of the change-over function in these two transitions are not equal, then the energy cannot be conserved precisely during a close encounter phase; the energy jumps to a different level. When the close encounters occur very often, then the energy begins to drift away from the initial value. 

By choosing larger values for the $n1$ and $n2$ parameter, the slope of the change-over function gets smaller and therefore the energy error gets smaller. But increasing the values of $n1$ or $n2$ also means that the Bulirsch-Stoer phase becomes longer and that larger close encounter groups are formed, which greatly impacts the performance of the code.

The differences in the energy conservation between GENGA and Mercury comes mostly from a different definition of the n2 parameter. In the test shown in Figure\,\ref{fig:MercuryEnergyP32}, GENGA seems to conserve the energy in some cases better than Mercury. Both codes, GENGA and Mercury, conserve the energy better than pkdgrav2, because they directly evaluate all interactions between the bodies, while pkdgrav2 uses a tree code and hence has a considerably larger truncation error in the computation of the forces. We note that pkdgrav2 doesn't use the hybrid symplectic integrating scheme, but the SyMBA method described in \cite{DuncanLevisonLee98}. Therefore one cannot make a one-to-one comparison between these codes.

The high frequency oscillations on the order of a dynamical time seen in the energy are characteristic of a symplectic integrator and depend principally on the eccentricity of the bodies. However, the drift in the energy due to the close encounter errors, shows that the method is symplectic only in an approximate sense.  Using higher order symplectic schemes will not give better energy conservation during the close encounter phases, but can reduce the overall fluctuations in the energy.

\subsection{Comparison to Mercury and pkdgrav2}
\label{Compare}
\begin{figure}
\epsscale{1.25}
\plotone{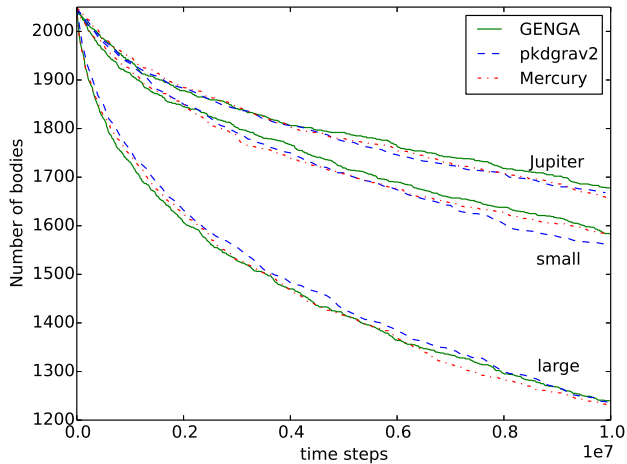}
\caption{Number of bodies as a function of time for a set of three different initial conditions, called ``small'', ``large'' 
and ``Jupiter''. The initial number of bodies is 2048 in all simulations, and decreases with time due to collisions or ejections from the system. The three codes GENGA, Mercury and pkdgrav2 show a very similar result.}
\label{fig:N} 
\end{figure}

We compare GENGA with Mercury\footnote{Mercury can be found at http://www.arm.ac.uk/~jec/} \citep{Chambers99} and pkdgrav2\footnote{An updated version of pkdgrav2 can be found at  http://hpcforge.org/projects/pkdgrav2/} \citep{Morishima+2010}.
We run a set of initial conditions with all three Codes. These sets of initial conditions are called ``small'', ``large'' and ``Jupiter''. They all consist of 2048 planetesimals distributed in a disk between 0.5 and 4 AU. The total mass of the planetesimals is set to 5 Earth masses in ``small'' and ``Jupiter'', and 50 Earth masses in the ``large'' set. In ``Jupiter'' we replaced one planetesimal from the ``small'' set with the planet Jupiter. In Figure\,\ref{fig:N} is shown the decrease in the number of bodies as a function of time for each of the three simulations. The number of bodies decreases either by collisional mergers of bodies or by ejections from the system. The three different codes produce a very similar result. One has to note that in a parallel code, the result of different runs of the same initial conditions can vary, due to a different execution order of the parallel parts, and due to different rounding errors.

In performing the simulations with Mercury, we respected the bug report from \cite{TorresAnderson08}.

\begin{figure}
\epsscale{1.25}
\plotone{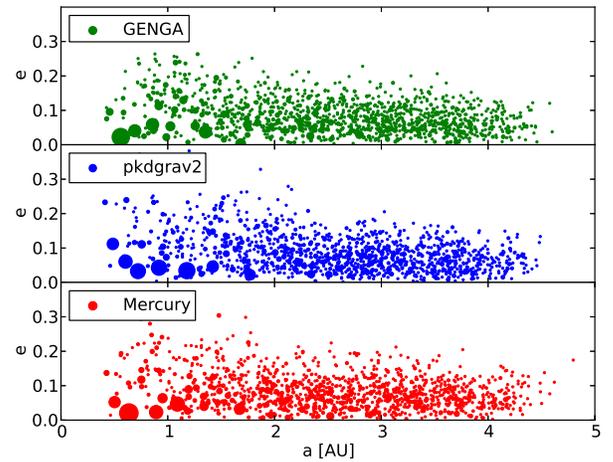}
\caption{Comparison between GENGA, pkdgrav2 and Mercury for the ``large'' simulation after 164,000 yr. The plot shows the semi-major axis vs. the eccentricity, while the size of the points represents the masses of the planetesimals. The three codes show a very similar result, but individual bodies are not comparable between the codes.}
\label{fig:aeLarge} 
\end{figure}

\begin{figure}
\epsscale{1.25}
\plotone{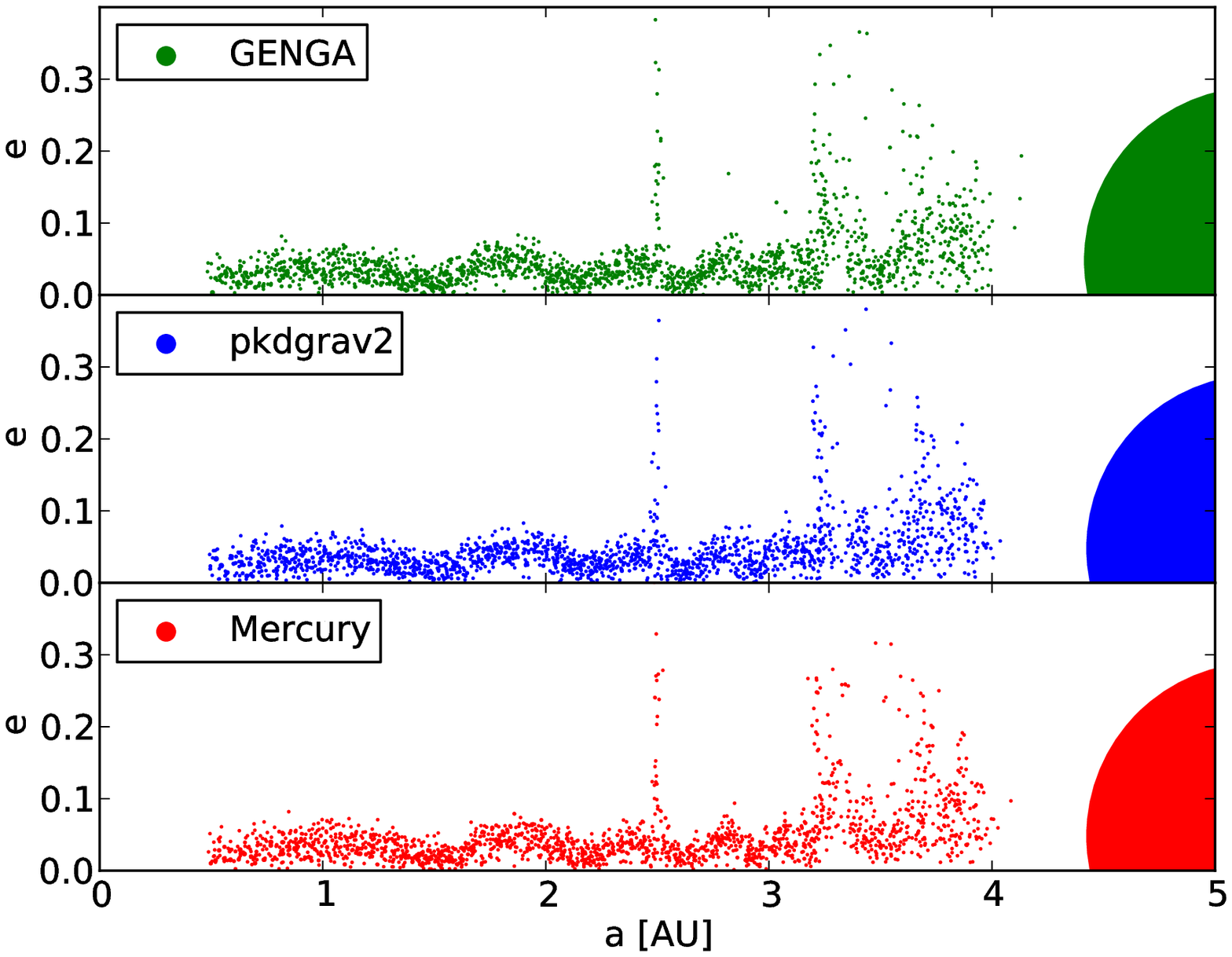}
\caption{Comparison between GENGA, pkdgrav2 and Mercury for the ``Jupiter'' simulation after 57,000 yr. The plot shows the semi-major axis vs. the eccentricity, while the size of the points represents the masses of the planetesimals. The three codes show a very similar result, but positions of individual bodies are not comparable between the codes.}
\label{fig:aeJupiter} 
\end{figure}

In Figure\,\ref{fig:aeLarge} is shown the semi-major axis versus the eccentricity after 164,000 yr for the ``large'' simulation, performed with the three different codes. Qualitatively the general results look very similar, but the individual planetesimals do not agree perfectly due to the chaotic nature of the $N$-body problem, where different rounding errors and/or execution order result in visible differences after many dynamical times.
In Figure\,\ref{fig:aeJupiter} is shown the same plot for the ``Jupiter'' simulation after 164,000 yr. Again the codes produce a very similar result. One can clearly see the 3:1 and 2:1 mean motion resonances between the planetesimal and Jupiter. The wave in the a-e plane appears because initially the planetesimals and Jupiter are not in the same orbital plane. After 200,000 yr this wave disappears.


\subsection{Performance}

\subsubsection{Performance comparison to Mercury and pkdgrav2}

\begin{figure}
\epsscale{1.25}
\plotone{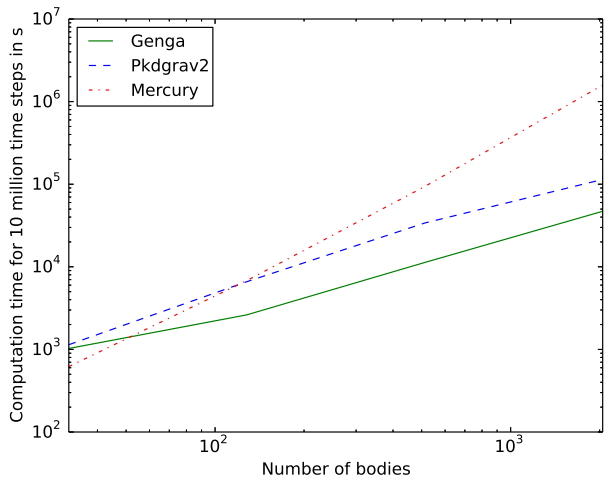}
\caption{Performance of Mercury, GENGA and pkdgrav2 for a set of four simulations with 32, 128, 512 and 2048 planetesimals, with each  a total mass of 5 Earth masses, distributed between 0.5 and 4 AU. With a high number of bodies, GENGA is up to four times faster than pkdgrav2 and up to 40 times faster than Mercury. At a low number of bodies, GENGA is slower than Mercury because of the slower clock rate of the GPU and because of the kernel overheads.}
\label{fig:CompareTime2} 
\end{figure}

In Figure\,\ref{fig:CompareTime2} is shown the performance of the three codes for the ``small'' simulation, described in the previous section and three additional simulations with 32, 128 and 512 planetesimals, also all with a total planetesimal mass of 5 Earth masses.

GENGA is up to four times faster than pkdgrav2, which uses a tree code to reduce the order of operations, and up to 40 times faster than Mercury. With only 32 bodies, the performance of GENGA is limited by kernel launch overheads and cannot benefit from the parallelization. The lower clockrate of the GPU and higher latency per instruction compared to a CPU core are also important factors here. It should be noted that while pkdgrav2 is a parallel code, it can only effectively make use of the number of cores on a single socket due to the very short execution time of a single time-step for this small number of particles. Communication latencies to other computing nodes in current HPC systems are simply too high to allow any speedup by distributing such a small number of bodies across the system.

\subsubsection{Performance of the Main Simulation Mode Kernels}
\begin{figure}
\epsscale{1.25}
\plotone{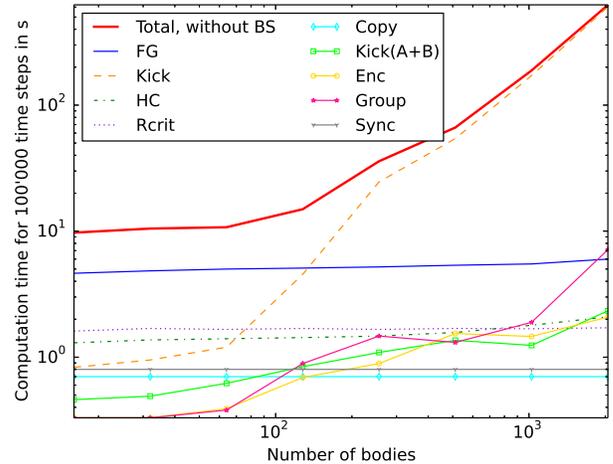}
\caption{Performance of the different kernels as a function of the number of bodies. The time of close encounter integration is not included in this plot, because this time depends strongly on the initial condition. For a large number of bodies, the kick kernel clearly dominates the execution time, while at a small number of bodies, the Keplerian drift is the most expensive. At a small number of bodies, the summation of all the kernel overheads plays an important role. This timing was done on an NVIDIA GTX 680 card.}
\label{fig:Timing} 
\end{figure}

The performance of the main kernels is shown in Figure\,\ref{fig:Timing} as a function of the number of bodies. The integrated system has a total mass of 5 Earth masses, distributed in a disk between 0.5 and 4 AU. In the plot we do not show the time needed for the Bulirsch-Stoer integration of the close encounter pairs, because this time depends strongly on the initial conditions. Particularly in the beginning of a simulation, the Bulirsch-Stoer integration can easily take ten times more execution time than the rest of the kernels. In a later phase of the integration, the Bulirsch-Stoer phase can vanish completely. The performance of a full simulation is shown later in Section\,\ref{Compare}.

At a small number of bodies, the FG kernel dominates, followed by the Rcrit and HC kernels, but also the summation of all the kernel overheads, data transfer and synchronization are important. At a large number of bodies, the kick kernel dominates because of the $N^2$ dependence. The group-, encounter- and kick(A+B) kernel also become more important at large $N$, but this depends also on the initial conditions and the chosen close encounter parameters. All the kernels with a linear dependence on the number of the bodies show an almost constant line in the performance plot. The reason for this is that these kernels do not manage to use the full GPU computing resources, but beyond a certain $N$ the GPU will be fully occupied and the execution time of these kernels will grow linearly with $N$, as expected.

\subsubsection{A Performance Comparison between Different GPU Cards}

\begin{figure}
\epsscale{1.25}
\plotone{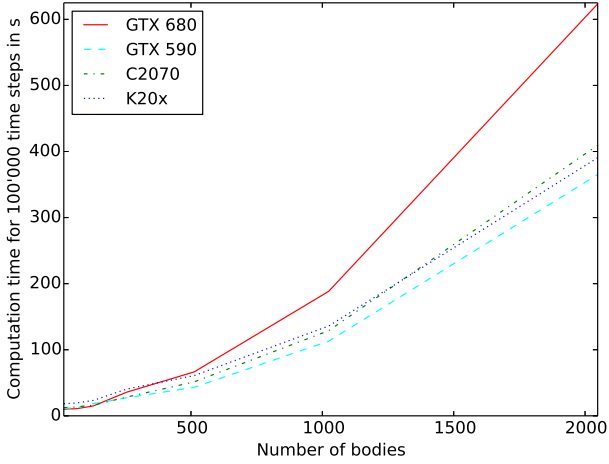}
\caption{Comparison of the performance on different GPU cards. The GTX 680 and GTX 590, the C2070 on Eiger at CSCS and the K20x on Todi at CSCS.}
\label{fig:Timing2} 
\end{figure}

In Figure\,\ref{fig:Timing2} is shown a comparison between four different GPUs, the GTX 680, GTX 590, C2070 and the K20x cards. 
Using a small number of bodies, the GTX 680 and the GTX 590 are the fastest ones, while the C2070 and the K20x show much more overhead time. For a large number of bodies the GTX 590 is the fastest, because the code was developed on this card and is optimized for this architecture. The new K20x has a much higher number of cores than all the others but it will need a different design of the kick kernel to be more efficient.

\subsubsection{Performance of the Test Particle Mode}
\begin{figure}
\epsscale{1.25}
\plotone{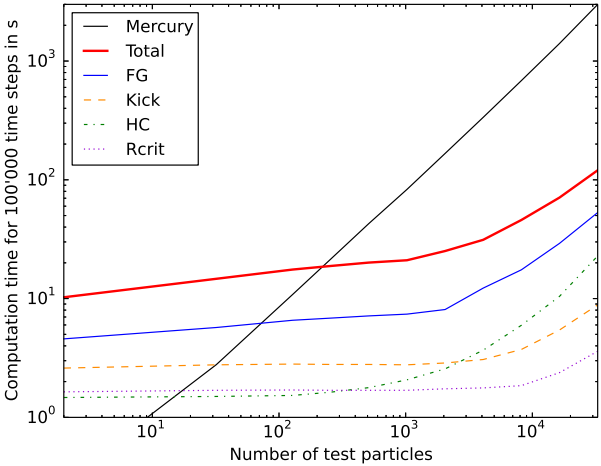}
\caption{Performance of the main kernels as a function of the number of test particles in a simulation with three massive bodies, tested on a GTX 680. Since the order of the kick kernel is only linear, its contribution is less important than the more complicated FG kernel. All of the kernel execution times grow after a few thousand particles because at this point the GPU is fully occupied.}
\label{fig:TP_Timing} 
\end{figure}

In Figure\,\ref{fig:TP_Timing} is shown the performance of the most important kernels in the test particle mode for a simulation containing three massive bodies and a variable number of test particles. All kernels show a similar curve, where the FG kernel is the most expensive one, followed by the kick kernel. The execution time of all kernels begin to grow after a few thousand test particles, because then the GPU is fully occupied. For a large number of test particles the HC kernel gets more expensive than the kick kernel. This can be improved by computing the total momentum of the system multiple times in different thread blocks, which simplifies the distribution of the momentum to the test particles.

\begin{figure}
\epsscale{1.25}
\plotone{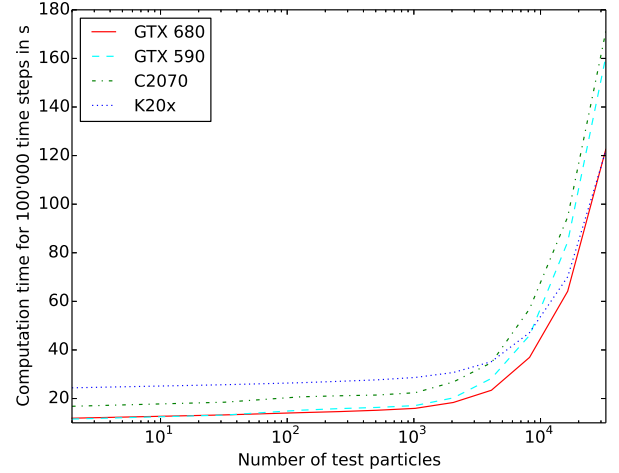}
\caption{Test particle mode performance comparison of different GPU cards. The GTX 680 and GTX 590, the C2070 on Eiger at CSCS and the K20x on Todi at CSCS. For a large number of particles the K20x is the fastest due to its larger core count, while at a low number of particles it becomes the slowest due to a higher overhead time.}
\label{fig:TP_Timing2} 
\end{figure}

In Figure\,\ref{fig:TP_Timing2} is shown a comparison between different GPU cards. With a small number of test particles, the GTX 680 and the GTX 590 are the fastest ones, while the C2070 and the K20x have much more overhead, because the integration is dominated by the complicated FG kernel. Only at a high number of test particles the K20x can benefit from the higher number of threads, and beats all the other cards.

\subsubsection{The Performance of the Multi-Simulation Mode}
\begin{figure}
\epsscale{1.25}
\plotone{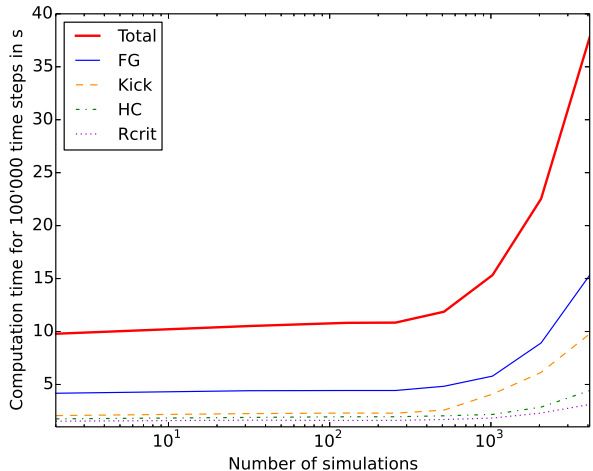}
\caption{Performance of the main kernels as a function of the number of three-body simulations, tested on a GTX 680. Most expensive is the FG kernel, followed by the kick kernel. Since all kernels have a linear dependence of the number of bodies, the execution time is almost constant and begins to grow when the GPU is fully occupied.}
\label{fig:M_Timing} 
\end{figure}

To test the performance of the kernels we integrated a system with three bodies many times in parallel. The performance as a function of the number of simulations is shown in Figure\,\ref{fig:M_Timing} for the most important kernels. In this test no close encounters appear.
In the multi-simulation mode, the computation time of the kernels have a nearly linear dependency on the number of simulations. Only in the HC and kick kernels are additional threads required at the boundaries of blocks. One can see clearly an increase in time going from 256 simulations to 512. In this regime the kernels are using the full GPU, and by using more simulations, some operations become serialized. The most expensive kernel is the FG, followed by the kick kernel. 

\begin{figure}
\epsscale{1.25}
\plotone{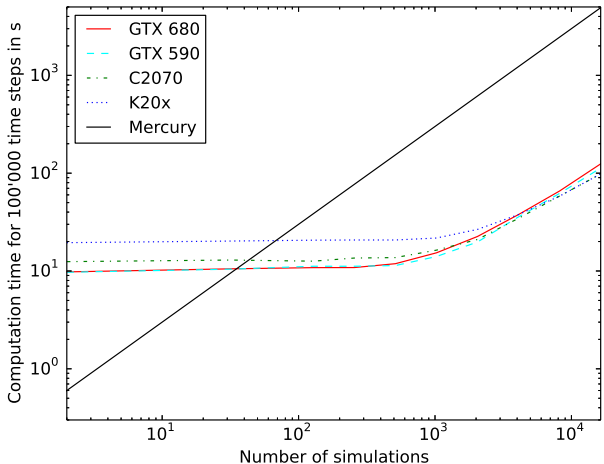}
\caption{Multi-simulation mode performance comparison between different GPU cards. The GTX 680 and GTX 590, the C2070 on Eiger at CSCS and the K20x on Todi at CSCS. Also shown is the execution time that Mercury needs to integrate all the systems on one CPU.}
\label{fig:M_Timing2} 
\end{figure}

In Figure\,\ref{fig:M_Timing2} is shown a performance comparison of different GPU cards. It shows also the time that Mercury needs to integrate the simulations on a 2.8 GHz Intel Xeon CPU core. Simulating only a small number of simulations is clearly faster on a CPU, but by using more than 30 simulations, the GPU becomes more efficient. Even with 16384 simulations the slope of the GPU performance is still smaller than that of the CPU.

It is very interesting to compare the different cards to each other. The K20x has much more cores than the C2070, which means it starts to serialize operations at a larger number of simulations. But since the K20x card shows a higher overhead time, it cannot benefit from the bigger number of cores, at least in the regime with less than 8192 simulations. At 8192 and at 16384 simulations the difference between the two cards is very small.
The GTX680 shows the least overhead time at a low number of simulations, but with many simulations, the slope is the steepest one.

\section{Limitations of the Hybrid Symplectic Integrator}
\label{Limits}
\begin{figure}
\epsscale{1.25}
\plotone{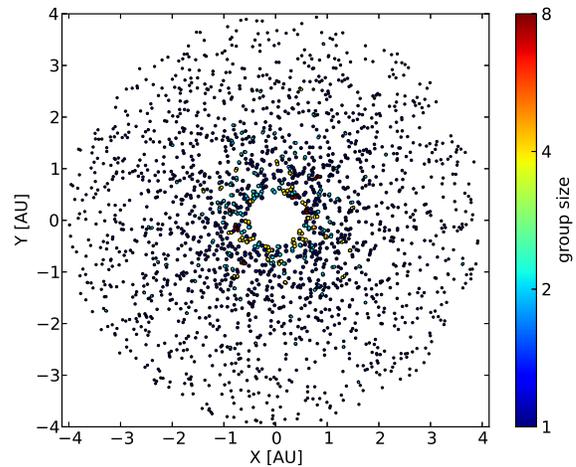}
\caption{Close encounter groups for a set of 2048 planetesimals with close encounter parameters $n1 = 3$, $n2 = 0.4$ and a time-step of 6 days. The color of the close encounter groups represents the number of bodies involved, while the size of the points correspond to their critical radius. The biggest close encounter groups consist of eight members, while the most of them are simply a close encounter pair.}
\label{fig:cluster3} 
\end{figure}

\begin{figure}
\epsscale{1.25}
\plotone{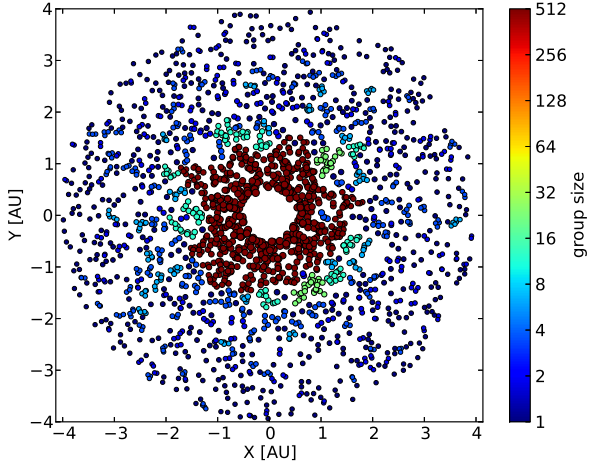}
\caption{As in the previous figure, the close encounter groups are shown, but this time for the parameters $n1 = 3$ and $n2 = 1.5$. All bodies in center of the disk are in a close encounter with enough neighbors so that all of them become chained together forming one big close encounter group. In the outer part of the disk, some smaller close encounter groups also occur.}
\label{fig:cluster}
\end{figure}

The main reason for limiting the number of bodies in the current version of GENGA to 2048 is the velocity dependence of the change-over function. Since the orbital velocity doesn't depend on the mass of the body, and since the critical radius of a small planetesimal is usually dominated by the $n2$ condition, a higher number density of planetesimals will cause a much higher number of ``close'' encounters. However, these are only identified as close encounters due to the requirement of the hybrid integrator that a sufficient number of steps be taken through the change-over function.  
In the central part of the disk, where the critical radius is typically the largest, the close encounters can be chained together and form very large groups. A close encounter chain can occur if we have for example two close encounter pairs $A-B$ and $B-C$. Then all three bodies $A$, $B$ and $C$ have to be treated as one close encounter group even if we do not have a direct close encounter between the bodies $A$ and $C$. In Figure\,\ref{fig:cluster3} and \ref{fig:cluster} are shown two examples for two sets of close encounter parameters. In Figure\,\ref{fig:cluster3} the biggest groups consists of 8 members, with most groups having just two members. In Figure\,\ref{fig:cluster}, all bodies of the inner part of the disk are chained together and form one big close encounter group.
We note that in the Mercury code the indirect close encounter pairs are not computed in the direct integration part. We think that our approach is more accurate in the sense of energy conservation, even if it is not that efficient in large close encounter groups. In future versions of GENGA this will probably change.

Combining this result with the energy conservation depending on the $n1$ and $n2$ parameters, described in Section\,\ref{Energy}, explains why the integration of more than 2048 bodies will be very inefficient with the current scheme. The problem could be solved by defining a different change-over mechanism.

\begin{figure}
\epsscale{1.25}
\plotone{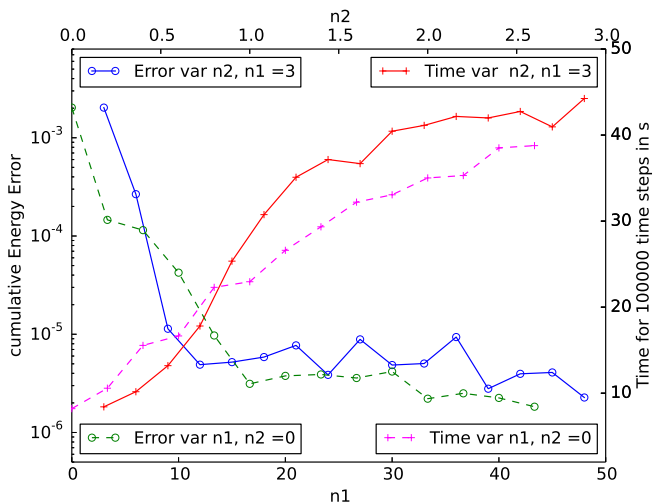}
\caption{Shown is the cumulative energy error and the execution time as a function of the close encounter parameters $n1$ and $n2$. The solid lines refers to a fixed value of $n1 = 3$ and a variable value of $n2$ drawn at the top axis of the diagram. The dashed lines refers to a variable value of $n1$ drawn at the bottom axis of the diagram and a fixed value of $n2 = 0$. The lines marked with a circle show the cumulative energy error drawn at the left axis, while the lines marked with a plus show the execution time drawn at the right axis. Increasing the values of $n1$ or $n2$ leads to a better energy conservation and a longer execution time.
}
\label{fig:ErrorTime} 
\end{figure}

In Figure\,\ref{fig:ErrorTime} we show the cumulative energy error, compared to the execution time of the simulation as a function of the $n1$ and $n2$ parameters, for one of the badly behaving simulations shown in Figure\,\ref{fig:MercuryEnergyP32}. Increasing the values of $n1$ or $n2$ improves the energy conservation, but requires also a longer execution time, due to more and longer close encounter phases. To achieve the same quality of energy conservation as the main part of the simulations in Figure\,\ref{fig:MercuryEnergyP32}, one should at least choose $n1 = 15$ or $n2 = 0.7$.

\subsection{Fixed Time Step Limit}
Another limit in the performance of the current integrator is due to the fixed time step in the symplectic integrator. The time step must be set according to the orbit of the innermost body. For bodies in the outer part of the system, a much longer time step would be sufficient for a comparable energy conservation, it would be more efficient to use an individual and adaptive time step integrator in the sense of \cite{PretoTremaine99} or \cite{MikkolaTanikawa99}. To include an individual and adaptive time step in GENGA is planned for future versions.

\section{Conclusions}
We presented the implementation of GENGA, a hybrid symplectic integrator designed and optimized for planetary system simulations. GENGA supports three simulation modes: Integration of up to 2048 massive bodies, integration with up to a million test particles, or parallel integration of a large number of individual planetary systems. We presented a detailed performance analysis of the code showing that at a large number of bodies, GENGA is up to 30 times faster than the Mercury code. At a very small number of bodies, GENGA is slower than Mercury due to GPU kernel overhead time and memory transfer between GPU and CPU.
We compared the results of GENGA to Mercury and pkdgrav2 and found a very similar qualitative behavior of planetary systems between the codes. We showed that the energy conservation of GENGA is better than Mercury and much better that pkdgrav2.
We presented the limitations of the current integration scheme and pointed out that future versions of GENGA should include an individual time stepping algorithm and a different changeover mechanism.

GENGA expands the second order hybrid symplectic integration scheme to fourth and sixth order, and has successfully been used with the test particle and multi-simulation mode to analyze the stability of exoplanetary systems \citep{Elser+13}. GENGA is available as open source code from https://bitbucket.org/sigrimm/genga.

\acknowledgments

We want to thank Sebastian Elser and Volker Hoffmann very much for their help on testing the code and improving its functionality. We thank Doug Potter for his technical support and for interesting discussions concerning GPUs.

GENGA was developed and tested on the GPU Tasna cluster which was purchased as part of the HP2C project ``Computational Cosmology on the Petascale''. We also used the zbox4 cluster at the University of Z\"urich and the Todi and Eiger systems at CSCS.

\appendix

\section{The Close Encounter Group Finding Algorithm}
\label{appendix}

\label{algorithm}

In the first step, all close encounter pairs are loaded into two arrays called P1 and P2, and an additional array LINK, with $N$ entries, is created and initialized with its own index (a link to itself). Index $e$ runs over all encounter pairs ($e \in$ EDGE), and index $i$ runs over all
particles ($i \in 1,...,N$). The object is to have for each group (an equivalence class defined through the list of encounter pairs) a unique index which can be found by following the links in the array LINK until $i == {\rm LINK}[i]$. This index is then the lowest index of the particles in the encounter group.
\[
{\rm LINK}[{\rm P1}[e]] := \min({\rm LINK}[{\rm P1}[e]],{\rm LINK}[{\rm P2}[e]])
\]
and
\[
{\rm LINK}[{\rm P2}[e]] := \min({\rm LINK}[{\rm P2}[e]],{\rm LINK}[{\rm P1}[e]])
\]
In the array LINK is now stored for each body the smaller index of the close encounter pair. 
One can add an extra optimization step which serves to reduce, by one, the number of links to be followed to find the smallest index over all the group members.
\[
{\rm LINK}[i] := {\rm LINK}[{\rm LINK}[i]].
\]
These steps are repeated until the array LINK remains unchanged.
The array LINK contains then for each body the smallest index over all the group members. In the algorithm the first two $min()$ operations need to be performed as $atomicMin()$ to prevent race conditions on accesses to the LINK array. We note that the LINK array is allocated in the high speed, but limited, shared memory making the $atomicMin()$ operations relatively efficient.

The next step is to transform the smallest index of a group into a consecutive group index, and to write the members of the groups line by line into a matrix. The sizes of the groups are written into a different array.

\subsection{Many Close Encounter Pairs}
For more than 512 close encounter pairs, the algorithm described before would use too much shared memory, and we have to split up the group search into several blocks. This means that the different blocks may only find parts of the groups, because other parts are found in different blocks. To link all the parts together we use a fusion kernel, which in principle uses the same algorithm as described before, but the concrete implementation depends strongly on the number of close encounter pairs. We implemented five different versions of the fusion kernel, each one can be used for a specific number of close encounter pairs and number of bodies. The different version are summarized in Table\,\ref{tab:fusion}.

\begin{table}
\begin{center}
\begin{tabular}{|c|c|c|}
\hline
 & $N \le 1024$ & $N > 1024$ \\
\hline
$512 < N_{\rm pairs} \le 1024$ & {\bf fusion} & {\bf fusion2} \\
\hline
unrestricted & \multicolumn{2}{|c|}{\bf fusionB}\\
\hline
$N_{\rm pairs} > 2048$ & {\bf fusionA} & {\bf fusionA2} \\
\hline
\end{tabular}
\end{center}
\caption{The five different fusion kernels. The kernels fusion and fusion2 merge two different subsets of close encounter groups. The fusionA and fusionA2 kernels merge two different subsets from a tree of close encounter groups, while the fusionB kernel merges several subsets of close encounter groups in serial.}
\label{tab:fusion}
\end{table}

\begin{figure}
\epsscale{1.0}
\centering
\plotone{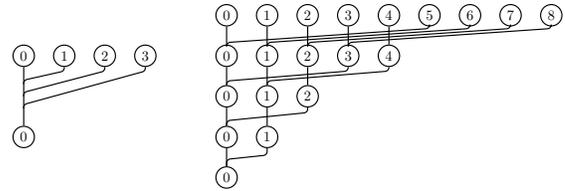}
\caption{Shown is the structure how subsets of close encounter groups are merged together. The left panel scheme shows how four subsets are merged with the fusionB kernel in a liner way. The right panel shows how more than four subsets are merged with the fusionA2 kernel by using a reduction formula. The last step in the right panel is performed with the fusionA kernel.
}
\label{fig:fusion} 
\end{figure}

If we have to split the group finding algorithm only in two subsets, then we can use the fusion or fusion2 kernel for merging together the subsets. If we split it up in three or four subsets, we use the fusionB kernel to merge the subsets in serial. If we split it up in more than four subsets we use the fusionA or fusionA2 kernels to merge the subsets with a tree structure. The last two subsets of the tree have to be merged with the fusion or fusion2 kernel. In Figure\,\ref{fig:fusion} is shown how the fusionA and fusionB kernels work.
The reason that we implemented different versions of the fusion kernel, depending on the number of close encounter pairs, is that the fusion operator can be called in each of the time steps, as demonstrated in Figure\,\ref{fig:cluster}. While it would be simpler to have a single kernel for this task, this significantly slows down simulations where there are many close encounters in each time step. This is also the typical scenario at the beginning of a planet formation simulation.

\bibliographystyle{apj}

\bibliography{mybib}

\begin{thebibliography}{}
\expandafter\ifx\csname natexlab\endcsname\relax\def\natexlab#1{#1}\fi

\bibitem[{{Aarseth} {et~al.}(1993){Aarseth}, {Lin}, \& {Palmer}}]{Aarseth+1993}
{Aarseth}, S.~J., {Lin}, D.~N.~C., \& {Palmer}, P.~L. 1993, \apj, 403, 351

\bibitem[{{Agnor} {et~al.}(1999){Agnor}, {Canup}, \& {Levison}}]{Agnor+1999}
{Agnor}, C.~B., {Canup}, R.~M., \& {Levison}, H.~F. 1999, \icarus, 142, 219

\bibitem[{Applegate {et~al.}(1985)Applegate, Douglas, G\"{u}rsel, Hunter,
  Seitz, \& Sussman}]{Applegate+1985}
Applegate, J.~H., Douglas, M.~R., G\"{u}rsel, Y., {et~al.} 1985, IEEE Trans.
  Comput., 34, 822

\bibitem[{{Asghari} {et~al.}(2004){Asghari}, {Broeg}, {Carone},
  {Casas-Miranda}, {Castro Palacio}, {Csillik}, {Dvorak}, {Freistetter},
  {Hadjivantsides}, {Hussmann}, {Khramova}, {Khristoforova}, {Khromova},
  {Kitiashivilli}, {Kozlowski}, {Laakso}, {Laczkowski}, {Lytvinenko}, {Miloni},
  {Morishima}, {Moro-Martin}, {Paksyutov}, {Pal}, {Patidar}, {Pe{\v c}nik},
  {Peles}, {Pyo}, {Quinn}, {Rodriguez}, {Romano}, {Saikia}, {Stadel}, {Thiel},
  {Todorovic}, {Veras}, {Vieira Neto}, {Vilagi}, {von Bloh}, {Zechner}, \&
  {Zhuchkova}}]{Asghari+2004}
{Asghari}, N., {Broeg}, C., {Carone}, L., {et~al.} 2004, \aap, 426, 353

\bibitem[{{B{\'e}dorf} \& {Portegies Zwart}(2012)}]{BedorfZwart2012}
{B{\'e}dorf}, J., \& {Portegies Zwart}, S. 2012, European Physical Journal
  Special Topics, 210, 201

\bibitem[{{Belleman} {et~al.}(2008){Belleman}, {B{\'e}dorf}, \& {Portegies
  Zwart}}]{Belleman+2007}
{Belleman}, R.~G., {B{\'e}dorf}, J., \& {Portegies Zwart}, S.~F. 2008, \na, 13,
  103

\bibitem[{{Capuzzo-Dolcetta} {et~al.}(2013){Capuzzo-Dolcetta}, {Spera}, \&
  {Punzo}}]{HiGPUs}
{Capuzzo-Dolcetta}, R., {Spera}, M., \& {Punzo}, D. 2013, Journal of
  Computational Physics, 236, 580

\bibitem[{{Chambers}(1999)}]{Chambers99}
{Chambers}, J.~E. 1999, \mnras, 304, 793

\bibitem[{{Chambers}(2001)}]{Chambers2001}
---. 2001, \icarus, 152, 205

\bibitem[{{Chambers}(2011)}]{ChambersBook2011}
---. 2011, {Terrestrial Planet Formation}, ed. S.~{Seager} (Tucson, Arizona:
  University of Arizona Press), 297--317

\bibitem[{{Chambers} \& {Wetherill}(1998)}]{ChambersWetherill1998}
{Chambers}, J.~E., \& {Wetherill}, G.~W. 1998, \icarus, 136, 304

\bibitem[{{Danby}(1988)}]{Danby88}
{Danby}, J.~M.~A. 1988, {Fundamentals of celestial mechanics} (Richmond,
  Virginia: Willmann-Bell)

\bibitem[{{de Souza Torres} \& {Anderson}(2008)}]{TorresAnderson08}
{de Souza Torres}, K., \& {Anderson}, D.~R. 2008, ArXiv e-prints,
  arXiv:0808.0483

\bibitem[{{Dindar} {et~al.}(2013){Dindar}, {Ford}, {Juric}, {Yeo}, {Gao},
  {Boley}, {Nelson}, \& {Peters}}]{SwarmNG}
{Dindar}, S., {Ford}, E.~B., {Juric}, M., {et~al.} 2013, \na, 23, 6

\bibitem[{{Duncan} {et~al.}(1998){Duncan}, {Levison}, \&
  {Lee}}]{DuncanLevisonLee98}
{Duncan}, M.~J., {Levison}, H.~F., \& {Lee}, M.~H. 1998, \aj, 116, 2067

\bibitem[{{Elser} {et~al.}(2013){Elser}, {Grimm}, \& {Stadel}}]{Elser+13}
{Elser}, S., {Grimm}, S.~L., \& {Stadel}, J.~G. 2013, \mnras, 433, 2194

\bibitem[{{Elser} {et~al.}(2012){Elser}, {Meyer}, \& {Moore}}]{Elser+2012}
{Elser}, S., {Meyer}, M.~R., \& {Moore}, B. 2012, \icarus, 221, 859

\bibitem[{{Gaburov} {et~al.}(2009){Gaburov}, {Harfst}, \& {Portegies
  Zwart}}]{Sapporo}
{Gaburov}, E., {Harfst}, S., \& {Portegies Zwart}, S. 2009, \na, 14, 630

\bibitem[{{Gladman} \& {Coffey}(2009)}]{GladmanCoffey2009}
{Gladman}, B., \& {Coffey}, J. 2009, Meteoritics and Planetary Science, 44, 285

\bibitem[{{Gladman} {et~al.}(2005){Gladman}, {Dones}, {Levison}, \&
  {Burns}}]{Gladman+2005}
{Gladman}, B., {Dones}, L., {Levison}, H.~F., \& {Burns}, J.~A. 2005,
  Astrobiology, 5, 483

\bibitem[{{Gladman} {et~al.}(1991){Gladman}, {Duncan}, \&
  {Candy}}]{Gladman+1990}
{Gladman}, B., {Duncan}, M., \& {Candy}, J. 1991, Celestial Mechanics and
  Dynamical Astronomy, 52, 221

\bibitem[{{Gladman} {et~al.}(1996){Gladman}, {Burns}, {Duncan}, {Lee}, \&
  {Levison}}]{Gladman+1996}
{Gladman}, B.~J., {Burns}, J.~A., {Duncan}, M., {Lee}, P., \& {Levison}, H.~F.
  1996, Science, 271, 1387

\bibitem[{Hairer {et~al.}(2011)Hairer, N{\o}rsett, \& Wanner}]{Hairer}
Hairer, E., N{\o}rsett, S., \& Wanner, G. 2011, Solving Ordinary Differential
  Equations I: Nonstiff Problems, Springer Series in Computational Mathematics
  (Berlin: Springer)

\bibitem[{{Hamada} \& {Iitaka}(2007)}]{HamadaIitaka2007}
{Hamada}, T., \& {Iitaka}, T. 2007, ArXiv Astrophysics e-prints,
  astro-ph/0703100

\bibitem[{{Hanslmeier} \& {Dvorak}(1984)}]{HanslmeierDvorak84}
{Hanslmeier}, A., \& {Dvorak}, R. 1984, \aap, 132, 203

\bibitem[{Harris(2008)}]{Harris07}
Harris, M. 2008, {Optimizing Parallel Reduction in CUDA}, Tech. rep., nVidia

\bibitem[{Horowitz \& Sahni(1983)}]{HorowitzSahni}
Horowitz, E., \& Sahni, S. 1983, Fundamentals of data structures, Computer
  software engineering series (Rockville, Maryland: Computer Science Press)

\bibitem[{{Hut} \& {Makino}(1999)}]{Grape}
{Hut}, P., \& {Makino}, J. 1999, Science, 283, 501

\bibitem[{{Kokubo} \& {Ida}(2002)}]{KokuboIda2002}
{Kokubo}, E., \& {Ida}, S. 2002, \apj, 581, 666

\bibitem[{{Kokubo} {et~al.}(2006){Kokubo}, {Kominami}, \& {Ida}}]{Kokubo+2006}
{Kokubo}, E., {Kominami}, J., \& {Ida}, S. 2006, \apj, 642, 1131

\bibitem[{{Laskar}(1989)}]{Laskar1989}
{Laskar}, J. 1989, \nat, 338, 237

\bibitem[{{Laskar}(1996)}]{Laskar96}
---. 1996, Celestial Mechanics and Dynamical Astronomy, 64, 115

\bibitem[{Laskar(2013)}]{Laskar2012}
Laskar, J. 2013, in Progress in Mathematical Physics, Vol.~66, Chaos, ed.
  B.~Duplantier, S.~Nonnenmacher, \& V.~Rivasseau (Springer Basel), 239--270

\bibitem[{{Makino} \& {Aarseth}(1992)}]{MakinoAarseth92}
{Makino}, J., \& {Aarseth}, S.~J. 1992, \pasj, 44, 141

\bibitem[{{Menou} \& {Tabachnik}(2003)}]{MenouTabachnik2003}
{Menou}, K., \& {Tabachnik}, S. 2003, \apj, 583, 473

\bibitem[{{Mikkola} \& {Tanikawa}(1999)}]{MikkolaTanikawa99}
{Mikkola}, S., \& {Tanikawa}, K. 1999, Celestial Mechanics and Dynamical
  Astronomy, 74, 287

\bibitem[{{Morishima} {et~al.}(2010){Morishima}, {Stadel}, \&
  {Moore}}]{Morishima+2010}
{Morishima}, R., {Stadel}, J., \& {Moore}, B. 2010, \icarus, 207, 517

\bibitem[{{Nitadori} \& {Aarseth}(2012)}]{NBODY6}
{Nitadori}, K., \& {Aarseth}, S.~J. 2012, \mnras, 424, 545

\bibitem[{{Nitadori} \& {Makino}(2008)}]{NitadoriMakino08}
{Nitadori}, K., \& {Makino}, J. 2008, \na, 13, 498

\bibitem[{Nyland {et~al.}(2007)Nyland, Harris, \& Prins}]{Gems3}
Nyland, L., Harris, M., \& Prins, J. 2007, Fast N-Body Simulation with CUDA

\bibitem[{{O'Brien} {et~al.}(2006){O'Brien}, {Morbidelli}, \&
  {Levison}}]{Obrien+06}
{O'Brien}, D.~P., {Morbidelli}, A., \& {Levison}, H.~F. 2006, \icarus, 184, 39

\bibitem[{{Portegies Zwart} {et~al.}(2007){Portegies Zwart}, {Belleman}, \&
  {Geldof}}]{PortegiesZwart+2007}
{Portegies Zwart}, S.~F., {Belleman}, R.~G., \& {Geldof}, P.~M. 2007, \na, 12,
  641

\bibitem[{{Preto} \& {Tremaine}(1999)}]{PretoTremaine99}
{Preto}, M., \& {Tremaine}, S. 1999, \aj, 118, 2532

\bibitem[{{Quinn} {et~al.}(1991){Quinn}, {Tremaine}, \& {Duncan}}]{Quinn91}
{Quinn}, T.~R., {Tremaine}, S., \& {Duncan}, M. 1991, \aj, 101, 2287

\bibitem[{{Raymond} \& {Barnes}(2005)}]{RaymondBarnes2005}
{Raymond}, S.~N., \& {Barnes}, R. 2005, \apj, 619, 549

\bibitem[{{Raymond} {et~al.}(2006){Raymond}, {Quinn}, \& {Lunine}}]{Raymond+06}
{Raymond}, S.~N., {Quinn}, T., \& {Lunine}, J.~I. 2006, \icarus, 183, 265

\bibitem[{{Reyes-Ruiz} {et~al.}(2012){Reyes-Ruiz}, {Chavez}, {Aceves},
  {Hernandez}, {Vazquez}, \& {Nu{\~n}ez}}]{ReyesRuiz+2012}
{Reyes-Ruiz}, M., {Chavez}, C.~E., {Aceves}, H., {et~al.} 2012, \icarus, 220,
  777

\bibitem[{{Richardson} {et~al.}(2000){Richardson}, {Quinn}, {Stadel}, \&
  {Lake}}]{Richardson+2000}
{Richardson}, D.~C., {Quinn}, T., {Stadel}, J., \& {Lake}, G. 2000, \icarus,
  143, 45

\bibitem[{{Saha} \& {Tremaine}(1992)}]{SahaTremaine1992}
{Saha}, P., \& {Tremaine}, S. 1992, \aj, 104, 1633

\bibitem[{{Stadel}(2001)}]{Stadel2011}
{Stadel}, J.~G. 2001, PhD thesis, UNIVERSITY OF WASHINGTON

\bibitem[{{Sussman} \& {Wisdom}(1988)}]{SussmannWisdom1988}
{Sussman}, G.~J., \& {Wisdom}, J. 1988, Science, 241, 433

\bibitem[{{Sussman} \& {Wisdom}(1992)}]{SussmaWisdom92}
---. 1992, Science, 257, 56

\bibitem[{{Wells} {et~al.}(2003){Wells}, {Armstrong}, \&
  {Gonzalez}}]{Wells+2003}
{Wells}, L.~E., {Armstrong}, J.~C., \& {Gonzalez}, G. 2003, \icarus, 162, 38

\bibitem[{Wilt(2013)}]{CudaHandbook}
Wilt, N. 2013, The CUDA Handbook: A Comprehensive Guide to GPU Programming
  (Pearson Education)

\bibitem[{{Wisdom} \& {Holman}(1991)}]{WisdomHolman91}
{Wisdom}, J., \& {Holman}, M. 1991, \aj, 102, 1528

\bibitem[{Yoshida(1990)}]{Yoshida90}
Yoshida, H. 1990, Physics Letters A, 150, 262

\end{thebibliography}

\end{document}